%% file: main.tex
\documentclass[conference]{IEEEtran}
\input{Macros.tex}
\begin{document}

\title{PEARL: Plausibly Deniable Flash Translation Layer using WOM coding}

\author{
\IEEEauthorblockN{
  Chen Chen, Anrin Chakraborti and Radu Sion
  }
\IEEEauthorblockA{
  Stony Brook University\\
  New York, USA\\
  }
}

\maketitle

\begin{abstract}
When adversaries are powerful enough to coerce users to reveal encryption keys, 
encryption alone becomes insufficient for data protection. Plausible deniability (PD) mechanisms resolve this by enabling users to hide the mere existence of sensitive data, often by providing plausible ``cover texts'' or ``public data volumes'' hosted on the same device. 

Unfortunately, with the increasing prevalence of (NAND) flash as a high-performance
cost-effective storage medium, PD becomes even more challenging in the presence of realistic adversaries who can usually access a device at {\em multiple points in time (``multi-snapshot'')}. This is because read/write operations to flash do not result in intuitive corresponding changes to the underlying device state.

The problem is further compounded by the fact that this behavior is mostly proprietary. For example, in a majority of commercially-available flash devices, an issued delete or overwrite operation from the upper layers almost certainly won't 
result in an actual immediate erase of the underlying flash cells. 

To address these challenges, we designed a new class of write-once memory (WOM) codes to store hidden bits in the same physical locations as other public bits. This is made possible by the inherent nature of NAND flash and the possibility of issuing multiple writes to target cells that have not previous been written to in existing pages. 

We designed \sysname{}, a general-purpose Flash Translation Layer (FTL) that allows users to plausibly deniably store hidden data in NAND flash devices. We implemented and evaluated \sysname{} on a widely used simulator FlashSim\cite{kim2009flashsim}. \sysname{} performs well on real-world workloads, comparably to non-PD baselines. {\em \sysname{} is the first system that achieves strong plausible deniability for NAND flash devices, secure against realistic multi-snapshot adversaries.}	
\end{abstract}

\section{Introduction}
\label{sec:intro}
As computers permeate aspects of daily life, individual users, government officials, and organizations store increasing amounts of sensitive and private data on personal
computers and mobile devices. While convenient, the ubiquitousness of computing devices that move
data with individuals poses increasing threats to privacy. There have been a number of high-profile cases
where a laptop or device with sensitive data is lost or stolen, leading to disclosure of sensitive information \cite{ex1,ex2,ex3,ex4,ex5}. Thus, protecting sensitive data is essential, as one can never anticipate when a
device may fall into the wrong hands.

To ensure sensitive data confidentiality, full disk encryption (FDE) is widely used.
However, considering adversaries who are empowered by law or otherwise to request encryption keys \cite{anderson1998steganographic,kennedy2000encryption,adv1,adv2,adv3},
the FDE alone may not be enough as it would be defeated by coercion of users into submitting the key or password to reveal confidential data.

Plausible deniability (PD) is a key security property that helps to protect sensitive data against the mentioned powerful adversaries. PD by definition makes it possible to claim that ``some information is not in possession [of the user] or some transactions have not taken place'' \cite{mcdonald1999stegfs}. 
In the context of secure storage, PD refers to the ability of a user to plausibly deny the existence of stored data even when an adversary has access to the storage medium.
It supplements the capability of encryption to protect sensitive data from powerful adversaries.

PD assurances are sometimes a matter of life and death~\cite{DBLP:conf/ndss/PetersGP15}. This has been demonstrated by numerous cases where information had to be transferred through checkpoints manned by hostile adversaries. One typical and prominent example involves the human rights group Network for Human Rights Documentation - Burma (ND-Burma). 
A large amount of data on human rights violations by the Burmese government was carried out of the country on mobile devices by ND-Burma activists, under threat of exposure at checkpoints and border crossings~\cite{case}.
Similarly, in 2012, a videographer smuggled evidence of human rights violations out of Syria by hiding a micro-SD card in a wound~\cite{syria}, again risking his life.

Several PD storage mechanisms were proposed \cite{mcdonald1999stegfs,anderson1998steganographic,pang2003stegfs,blass2014toward,chakraborti2017datalair,chang2018mobiceal,chen2019} for both file system and block device layers.
However, a strong assumption underpins all these existing solutions, mostly deriving from traditional magnetic media, namely a high level of transactional commitment from the underlying storage medium. Specifically, write and erase operations are assumed to be honored when issued. 

Needless to say, storage media such as NAND flash is wrapped in logic that prevents this to be the case. For example, most Flash Translation Layer (FTL) algorithms will likely prevent overwrites to touch underlying physical pages when issued and instead remap data elsewhere, only to later return and garbage collect such erased data if and when needed. 
This immediately breaks existing PD mechanisms built upon the assumption that the underlying device honors write/erase operations when requested. Stale data persisted on the underlying device (e.g., yet to be garbage collected pages) out of control of the PD logic then enables adversaries to easily infer the existence and most often location of hidden data \cite{jia2017deftl}. 

New media requires new PD logic. Further, arguably, this logic needs to be placed closer to the physical layer to securely handle the PD requirements while also providing life-cycle and efficiency-related elements such as wear leveling and encoding optimizations. 

NAND flash, arguably the most popular flash technology in modern production, stores data in an array of cells, each requiring a special ERASE operation before a write. Due to several addressing and packaging optimization reasons, almost always ERASE can only be performed at block level (containing many cells). As a result, even simple updates to data require a more complex set of steps which is implemented usually in an intermediary Flash Translation Layer (FTL) sitting between e.g., a file system and the underlying flash device. The FTL makes an excellent candidate \cite{jia2017deftl} for implementing protection functionality including PD logic. 

Two existing have considered PD tailored for NAND flash: DEFY \cite{DBLP:conf/ndss/PetersGP15}, and DEFTL \cite{jia2017deftl}. Unfortunately, neither is secure against practical adversaries which are almost always multi-snapshot \cite{chen2019}. Crossing a border twice, checking in airline luggage, living under an oppressive government with physical access to devices, leaving devices in untrusted places subject to ``hotel maid'' attacks, all these are instances of multi-snapshot opportunities for an adversary. Naturally, the security of a PD system should not break down completely (under reasonable user behavior) and should be resilient to such realistic externalities (hotel maids, border guards, airline checked luggage etc). Further, DEFY is compromised in the presence of capacity exhausting attacks \cite{jia2017deftl}.

{\it \sysname{} introduces the first PD scheme that achieves security against multi-snapshot adversaries on NAND flash devices.} This is made possible by re-purposing a new class of write-once memory (WOM) codes to naturally combine both public and hidden data together in one physical page, and managing the pages considering the nature of flash memory. \sysname{} is implemented as a general purpose FTL that, in addition to taking all necessary flash management duties, enables deniability of the existence of hidden data. 
It guarantees that the resulting state of a device with both public and hidden data is indistinguishable from a public-data only state. A number of key insights ground the design as follows. 

First, \sysname{} operates at a much finer encoding granularity compared with previous PD schemes.  
Existing work \cite{jia2017deftl,DBLP:conf/ndss/PetersGP15} store public and hidden data in different physical pages or even different flash blocks.
These systems require plausible reasons to explain away the existence of written pages containing hidden data (e.g. masquerading as ``random'' or ``free'' data). This problem is compounded by the nature of NAND flash and the realistic adversaries with multi-snapshot access. Hidden data may end up being relocated even in the absence of hidden updates, and adversaries can observe implausible modifications (e.g., to ``free'' space containing hidden data).
In contrast, \sysname{} uses the second write stage of a specially-designed WOM code to encode hidden data in a public cover. This makes such plausible reasons inherent -- all pages contain public data by design. 

Second, \sysname{} as an FTL smartly manages the mapping from both public and hidden data to physical pages and handles NAND-specific operations such as garbage collections considering the special nature of flash memory.
As a result, all physical layer changes can be plausibly explained by public data requests only, thus preventing multi-snapshot adversaries from detecting the existence of hidden data by comparing snapshots and analyzing physical activities on flash. 

We evaluated \sysname{} using a widely used simulator FlashSim \cite{kim2009flashsim}. The experimental results show \sysname{} is practically fast. It performs comparably to the non-PD baseline on real-world workloads.

%\section{Background and Related Work}
\section{Related Work}
\label{related}

%\subsection{Plausibly-Deniable Storage}
PD storage systems are designed to protect users against powerful adversaries (e.g., corrupt government officials) who can coerce users to give up the encryption key(s).
Generally speaking, a PD storage system allows the user to only reveal the key used to encrypt (non-sensitive) public data while claiming that no other data exists on the device. 

Steganographic file systems \cite{anderson1998steganographic,mcdonald1999stegfs,pang2003stegfs,DBLP:conf/ndss/PetersGP15} were firstly proposed to provide plausibly-deniable storage. 
% Steganographic file systems provide plausibly-deniable storage. 
They allowed users to store both sensitive (hidden) files and non-sensitive (public) files inside one file system and hide the existence of hidden files from adversaries.
To defend against single-snapshot adversaries, Anderson \etal \cite{anderson1998steganographic} explored the idea of steganographic 
file systems and proposed two ideas for hiding data.
% The first idea used a set of
% cover files and their linear combinations to reconstruct a hidden
% file. The ability to correctly compute the linear combination
% required to reconstruct a file was based on the knowledge of
% a password. This solution is not practical since the performance penalty is too
% high. Their second solution used a hash based scheme
% for storing files at locations determined by the hash of the
% filename. This solution required storing multiple copies of each file at different locations to prevent data loss. 
Later McDonald \etal \cite{mcdonald1999stegfs} implemented StegFS for
Linux on the basis of the solution proposed in \cite{anderson1998steganographic}.
Pang \etal \cite{pang2003stegfs} improved on the previous constructions by
avoiding hash collisions and provided more efficient storage.
In addition to these steganographic file systems against single-snapshot adversaries, Han \etal \cite{han2010multi} designed a multi-user steganographic file system (DRSteg) on shared storage.
% Dummy-Relocatable Steganographic file system (DRSteg) that allows multiple users to share the same hidden file. 
% They used the concept of run-time relocation of data to ensure deniability against a multi-snapshot adversary.
However, their solution does not scale well to practical scenarios as they attribute deniability to joint ownership of sensitive data.
Gasti \etal \cite{gasti2010deniable} proposed a deniable shared file system (DenFS) specifically for cloud storage. 
Its security depends on processing data temporarily on a client machine, and it is 
not straightforward to deploy DenFS for local storage.

On the other hand, disk encryption tools \cite{Truecrypt,rubberhose,skillen2013implementing,blass2014toward,chakraborti2017datalair,chang2018mobiceal,chen2019} were designed to support PD at block device level. 
They worked by often storing both hidden and public ``volumes'' on the same device while preventing adversaries to gain information about how many volumes the device actually contains. 
Truecrypt \cite{Truecrypt}, Rubberhose \cite{rubberhose} and Mobiflage \cite{skillen2013implementing} provided deniability against only single-snapshot adversaries. 
Blass \etal \cite{blass2014toward} implemented HIVE, the first PD solution against multi-snapshot adversaries at device level, using a write-only Oblivious RAM (ORAM) for mapping data from logical volumes to underlying devices and hiding access patterns for hidden data within requests to public data.
Later Chakraborti \etal \cite{chakraborti2017datalair} proposed DataLair with a more efficient write-only ORAM and improved the system performance.
Chang \etal \cite{chang2018mobiceal} proposed MobiCeal specifically for mobile devices. The idea is to use a dummy write mechanism to obfuscate writes to a hidden volume. Unfortunately the paper suffers from deniability compromises: the space occupied by dummy writes would be reclaimed while the space occupied by the hidden data would remain intact, thus enabling an attacker to detect the static hidden data. Chen \etal \cite{chen2019} introduced PD-DM, a locality-preserving PD solution that eliminated the randomness introduced by ORAM-based solutions and improved the system throughput especially on hard disk.

The above solutions required that the underlying devices honor write/erase operations atomically. Unfortunately in the case of flash this is simply not the case. Old data can linger on the device for years and attackers can easily unscrew the flash cover and read the FLASH chips directly with cheap off the shelf readers. Others have noted this too \cite{jia2017deftl} -- PD systems incorporating deniability in the upper layers (file system layer or block device layer) very often suffer from deniability compromises in the lower layers (flash memory). And unfortunately even systems such as Mobiflage and MobiCeal specifically designed for mobile devices do not address this essential vulnerability.

Special PD solutions are designed for NAND flash storage devices as well, considering its significant distinctive natures. 
DEFY~\cite{DBLP:conf/ndss/PetersGP15}
%is the only known file system that protects against a multi-snapshot adversary for local storage deployment. DEFY 
is a log structured file system for NAND flash devices that offers PD with a newly proposed secure deletion technology. 
It is based on WhisperYAFFS~\cite{whisper}, a log structured file system which provides full disk encryption for flash devices. However, as claimed in \cite{jia2017deftl}, DEFY will be compromised by making several attempts to exhaust the writing capacity. 
DEFTL \cite{jia2017deftl} instead incorporates deniability to the Flash Translation Layer (FTL) of flash-based block devices. Yet, it is against single-snapshot adversaries.

\section{NAND Flash}
%\subsection{NAND Flash and Flash Translation Layer}
\label{sec:nand}

%\paragraph{NAND Flash Memory}
NAND flash is a non-volatile solid-state storage medium. It is becoming increasingly popular due to its low power consumption and shock resistance now.
Unlike the traditional magnetic storage disk that stores data by magnetizing the ferromagnetic material on a disk, NAND flash stores data 
using only electronic circuits (floating-gates). Thus, NAND flash has its own characteristics \cite{grupp2009characterizing}:
% -- 
1) NAND flash supports efficient random accesses. 2) Read and write/program operations are performed in page units while erase operations are based on block units (usually larger than the page size by 64 or more times). 3) In addition to a data area, a page in NAND flash also contains a small spare OOB area which may be used for storing a variety of information such as the Error Correction Code (ECC) bytes, the logical page number and the page state. 4) An {\em erase} operation is required before writing in NAND flash. A floating-gate is charged during writing while only an erase can remove the charge from the gate. 5) NAND flash can withstand only a finite number of program-erase cycles (P/E cycles).

\subsection{Flash Translation Layer (FTL)}
To use NAND flash devices, we need either a file system specifically for raw NAND flash or a Flash Translation Layer (FTL) between the file system and the raw flash device. Some of the example NAND flash file systems that have been added to Linux kernel are UBIFS \cite{ubifs} and F2FS \cite{lee2015f2fs}. 
On the other hand, the FTL is an intermediate software layer between the host application (e.g. file systems) and NAND flash. It accepts logical requests from host and maps the logical addresses (LBAs) to physical addresses of the NAND flash.
% In commercial SSDs, the FTL usually exists as a firmware inside. But for some other products such as certain PCIe flash cards \cite{batwara2012leveraging}, it could be possible that the FTL runs on the host, using host CPU cycles and DRAM.

In addition to the logical-to-physical address mapping, a FTL is also responsible for some other necessary flash management duties such as wear leveling, garbage collection and so on. Wear leveling aims to smoothly distribute erases among blocks in the flash so that the blocks all reach their P/E cycle limit at the same time. Garbage collection is designed to efficiently reclaim pages that are no longer needed (i.e. invalid) in the device. Remembering that these pages cannot be simply erased at your leisure as they may be in blocks that still contain active data (i.e. valid). Instead, the FTL do the page recycle following these three steps: 1) adaptively select a victim block to be erased; 2) transparently move active data elsewhere; 3) erase the victim block.

According to how the logical-to-physical address mapping is performed, FTL schemes can be categorized into three groups: page-level FTLs, block-level FTLs and hybrid FTLs. The page-level FTL maps any logical page from the host to a physical page in the flash while the block-level FTL maps a whole logical block (containing multiple logical pages) to a physical block in flash. 
The hybrid FTL combines the page-level and block-level FTL by logically partitioning flash blocks into data blocks and log blocks. Data blocks are mapped with the block-level mapping while the log blocks are mapped using the page-level mapping scheme. Updates are written to log blocks, after which merge operations may happen to combine the active pages in data blocks and log blocks together as new data blocks.  \sysname{} deploys a page-level FTL based on DFTL\cite{gupta2009dftl}.

\subsection{Demand-based FTL (DFTL)} 
\label{section:DFTL}
\begin{figure}[t!] 
\centering
\includegraphics[width=8cm]{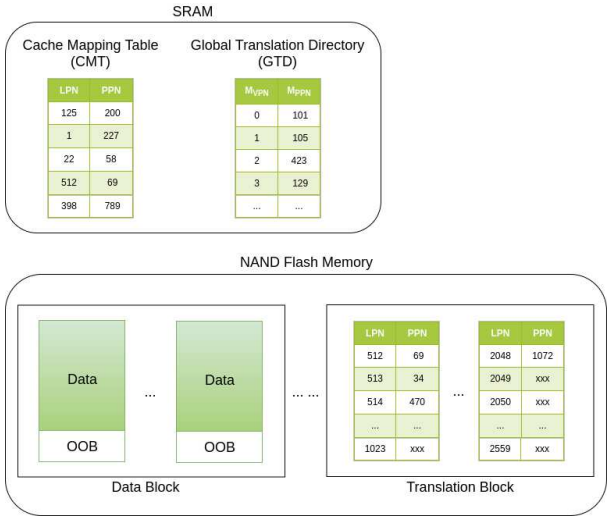}
\caption{\small The organization of DFTL. LPN is the Logical data Page Number, PPN is the Physical Page Number, $M_{VPN}$ is the Virtual Translation Page Number, $M_{PPN}$ is the Physical Translation Page Number.}
\label{fig:dftl}
\vspace{-0.3cm}    
\end{figure}

% \sysname~ is based on DFTL \cite{gupta2009dftl}. It 
DFTL is an efficient page-level FTL that avoids the inefficiency of hybrid FTLs and reduces the SRAM requirement for the page-level mapping. 
The page-level mapping table is stored in the flash memory and only a small amount of active mapping entries are cached in SRAM. A data structure called Global Translation Directory (GTD) is used to keep track of the whole mapping table scattered over the flash device. Figure \ref{fig:dftl} shows the organization of DFTL. 
% We choose DFTL as the base of our plausibly-deniable FTL since the page-level mapping is essential for the deniability, which will be detailed later in Section \ref{sec:sys}. 
% %Before introducing how \sysname~ is designed in detail,
% Here we firstly introduce how DFTL works.

% The page-level mapping table is packed into pages (translation pages) in the order of Logical data Page Numbers (LPNs) and stored in the flash memory. 
% A small amount of mapping entries are cached in SRAM as a Cached Mapping Table (CMT). A Global Translation Directory (GTD) is used to keep track of those translation pages scattered over the flash device. Figure \ref{fig:dftl} shows the organization of DFTL.

\paragraph{Logical-to-physical address translation}
% The address translation is the primary responsibility of any FTL. 
The address translation in DFTL is related to three data structures: the page-level mapping table, the Global Translation Directory (GTD) and the Cached Mapping Table (CMT). As shown in Figure \ref{fig:dftl}, the page-level mapping table is packed into pages (named as translation pages) in the order of Logical data Page Numbers (LPNs) and stored in translation blocks in the flash.
% , while the CMT and the GTD are both in the SRAM. 
The CMT stores the mapping entries (LPN-to-PPN) for those most recently accessed data pages and updates them using the segmented LRU array cache algorithm \cite{karedla1994caching}. The GTD maintains the physical page address information for all the translation pages. One translation page could store 512 mapping entries, if an address is represented in 4 bytes and the page size is 2KB. In this case, the first translation page with $M_{VPN}=0$ stores the mapping information for the first 512 logical pages and so forth, and the location of this translation page will be the first entry in the GTD. 
Both the CMT and the GTD are stored in the SRAM.

Once a logical request comes, the DFTL will first query the CMT for the mapping information. The request will be directly fulfilled if the mapping is found. Other wise, the DFTL fetches the mapping information from the flash into the CMT by the follow steps: 1) it checks the GTD for the physical location of the corresponding translation page; 2) it reads the translation page for the mapping and adds it into the CMT. A CMT eviction may happen during the above procedure. The evicted item needs to be written back only if it has been changed after loaded. This consists of 3 steps: 1) locate the corresponding translation page by consulting the GTD; 2) read the translation page and write it back to a new physical location with updated information. 3) update the corresponding GTD entry.
After the coming logical request is performed, the mapping information may be updated if necessary. Note that it will be always updated in CMT. The update to the translation pages on flash will only happen if a CMT eviction happens.

\paragraph{Page allocation and garbage collection}
In DFTL, data pages are written into data blocks whereas translation pages are written into translation blocks. DFTL maintains two blocks called Current Data Block and Current Translation Block for the page allocation. A free block will be chosen as the new Current Data Block or new Current Translation Block from a free block list when pages in either of the two blocks are used up. The garbage collector will choose the block with the least number of active pages as the victim to recycle. If the victim block is a translation block, DFTL copies the active translation pages to the Current Translation Block and update the GTD before erasing the victim block. Otherwise, if the victim is a data block, DFTL relocates the active data pages to the Current Data Block and update the corresponding mapping information in the CMT. 

\section{Model}
\label{sec:model}

% In this paper, we consider PD storage on NAND flash based devices. 
%Figure \ref{fig:model} shows the organization of a NAND flash storage device with an FTL supporting PD. 

% \todo{
% A user stores data on a device ... 
% the adversary ... 
% ... as a result the user wants to hide data ... 
% in existing systems it does so by having ...}

%\blue{

%
In a typical scenario, a user requires secure data storage for sensitive {\em hidden data} (which needs to be protected from powerful adversaries), and less sensitive {\em public data} (which do not require any special protection mechanisms). The adversary is coercive and can compel the user to hand over encryption keys etc. Under duress, the user may need to reveal keys to public data while denying the existence of the hidden data. An effective PD system should therefore not only hide the contents of the hidden data but also its very existence. 

\paragraph{Deployment}
{\sysname~} incorporates the PD functionality in the NAND flash FTL. 
Specifically, \sysname{} stores multiple logical block volumes on one physical flash device -- some of the volumes store hidden data while others store public data. W.l.o.g., for simplicity, we discuss here a design with only two volumes.  The data in the public and hidden volumes are encrypted with different encryption keys,  $K_{pub}$ and $K_{hid}$ respectively. The keys may be securely derived from user-generated passwords or other more secure mechanisms.

{\sysname~} can be used either in a public-only mode -- in which case the user can only access public data -- or in a public+hidden mode where the user can access both hidden and public data. To determine the mode of operation, the user provides appropriate passwords/keys at boot time (or when the device is plugged in after a reboot etc). For the public-only mode, the user provides $K_{pub}$; to access also hidden data both $k_{hid}$ and $K_{pub}$ are required. Note that under coercion, the user will reveal $K_{pub}$ to the adversary and operate in the public-only mode. As we will see, PEARL ensures that an adversary observing flash state 
does not gain a non-negligible advantage in detecting the existence of $K_{hid}$ or of any hidden data. 

When hidden data is stored on the device, {\sysname~} should be operated in the public+hidden mode since the system running in public-only mode (without the hidden key) may overwrite hidden data (e.g., during garbage collection). As discussed later, hidden data is relocated before an ERASE during garbage collection. Without the hidden key, {\sysname~} cannot re-encrypt and relocate this data to new locations. This is a common assumption for NAND flash PD solutions\cite{DBLP:conf/ndss/PetersGP15}.

\paragraph{Adversary}
% As in prior work\cite{anderson1998steganographic,mcdonald1999stegfs,blass2014toward,chakraborti2017datalair}, 
When defining a threat model, it is important to also consider any hardware-related characteristics that may result in adversarial advantages. The PD adversaries we consider come with the following assumptions:

\begin{itemize}
    \item Although adversaries can coerce users into giving up encryption keys, they are computationally bounded and ``rational'' -- they stop coercing users if no evidence of hidden data is observed. 
    
    \item Adversaries are aware of the underlying design of a PD system. In other words, the goal is not to provide security through obscurity. But at the same time, the mere presence of a PD system in the software stack will not serve as evidence that the user is hiding information. Ideally, once plausible deniability systems become efficient enough, they will be simply deployed in the standard OS codebase. 
    Therefore, a flash device with {\sysname~} will not be a red flag to the adversary.  
    % Note that with only one notable exception \anrin{can you cite INFUSE here?}, all existing PD systems make this assumption. 

    %they are not allowed to make assumptions about the presence of the hidden data based on merely the availability of a PD system. Thus, although adversaries are aware of the capability of devices to store hidden data, whether users choose to actively
    % use the hidden volume and 
    %store any hidden data in the flash devices is completely hidden from adversaries. In other words, the availability of a PD system will not be a red flag. 
    % Adversaries are aware of the capability of devices to store hidden data. However, what is hidden from the adversaries is whether users choose to actively use the hidden volume and store any hidden data in the flash devices. The availability of a PD system will not be a red flag as adversaries are not allowed to make assumptions about the presence of the hidden volume based on merely the availability of a PD system.
    %Ideally, a FTL supporting PD is supposed to be a widely used firmware. 
    
    \item Adversaries have "multi-snapshot" capabilities and can access the raw image\footnote{The raw image of a devices is not hard to acquire. For example, in many SSDs, this can be easily achieved by opening the covers and directly reading the memory chips with cheap off the shelf readers.} of a user's NAND flash device arbitrary number of times. Note that existing work on flash-based PD systems considers a weak "single-snapshot" adversary limited to only observing the flash memory {\em once} in its lifetime.
    
    %This is different from the adversaries in existing work whose ability to snapshot the device is limited to be a single time only. This is denoted as multi-snapshot adversaries, which are more realistic. 
    
    %Note that the  
    \item Adversaries can access the physical device only after it is unmounted or powered off \cite{blass2014toward} (these are commonly denoted as ``on-event'' adversaries). Thus, the running state of the device and the DRAM contents cannot be captured by the adversary. 
    Indeed, otherwise in the presence of an online adversary capable of monitoring user I/O and device state at runtime, arguably it would be close to impossible to provide strong plausible deniability.

    % Adversaries do not have access to the storage devices when they are actively used by users. Instead, 
\end{itemize}

Note that as is the case with all existing PD solutions, {\sysname~} does not protect against denial of service attacks. Specifically, the adversary can overwrite all hidden or public data on disk. 

\begin{figure}[t!] 
\vspace{-0.3cm}   
\centering
\includegraphics[width=5.5cm]{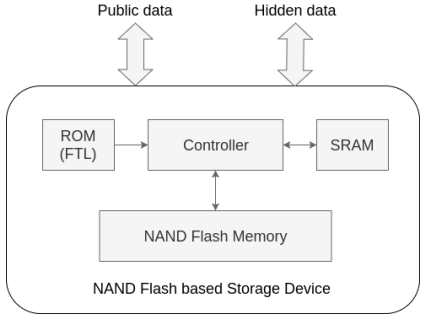}
\caption{The organization of a NAND flash storage device with an FTL supporting PD.}
\label{fig:model}
\vspace{-0.1cm}    
\end{figure}

\section{Hiding Data Using WOM Codes}
\label{sec:pdwom}

\sysname{} hides information by {\em modulating the written public data according to the data to be hidden}. As we will show, this is something that WOM codes can be re-purposed for. The end-result of hiding information is a device state that is indistinguishable from the case of a device that was simply writing data multiple times using a WOM code.
In this section, we show how it is indeed possible to design such a data encoding scheme by leveraging a special group of write-one-memory (WOM) codes.

\subsection{Overview}

\begin{figure}[t!] 
\centering
\subfigure[A WOM code allows multiple writes to the same physical locations by flipping some of the bits from 0 to 1. In this example, an initial write of 8 data bits results in setting 3 bits of ``1'' among 12 physical bits. Then, {\em later} in a second write, a {\em completely different set} of 8 bits of data can be written to the same locations by setting another 3 bits to 1. In the end, the 12 physical bits $010,000,100,001$ represent data $11,11,11,11$]{\label{fig:sub1} 
\includegraphics[width=.95\linewidth]{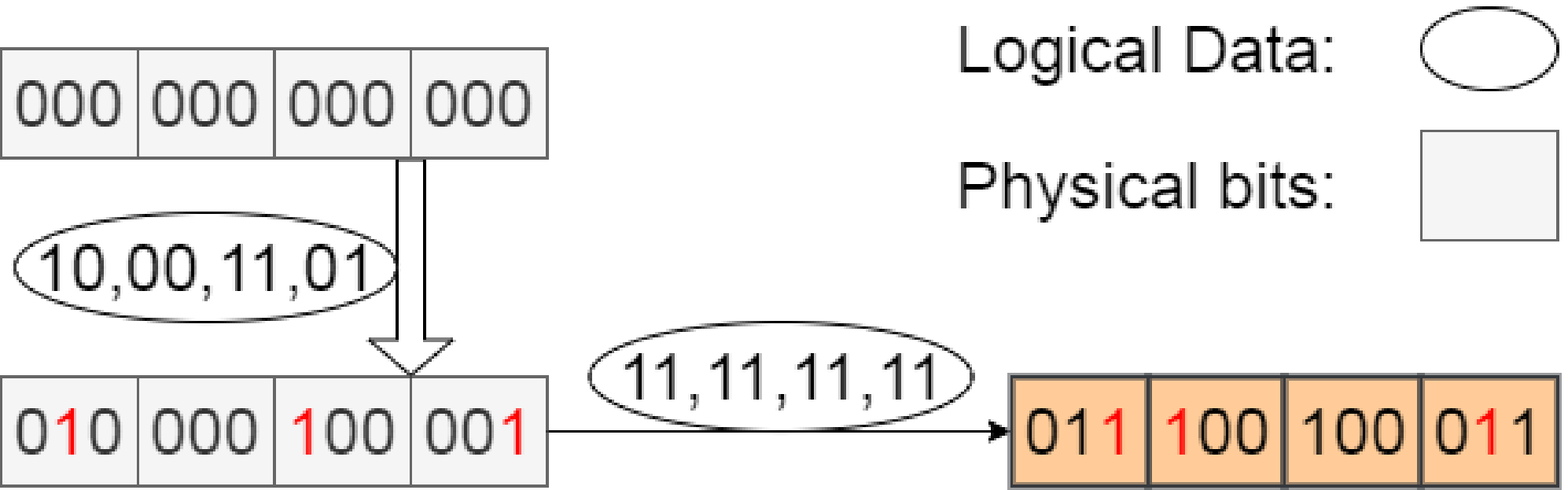}}
\subfigure[This is possible because the WOM code in the above example allows both $011$ and $100$ bit configurations (codewords) to represent data $11$ in the underlying device]{\label{fig:sub2} \includegraphics[width=.55\linewidth]{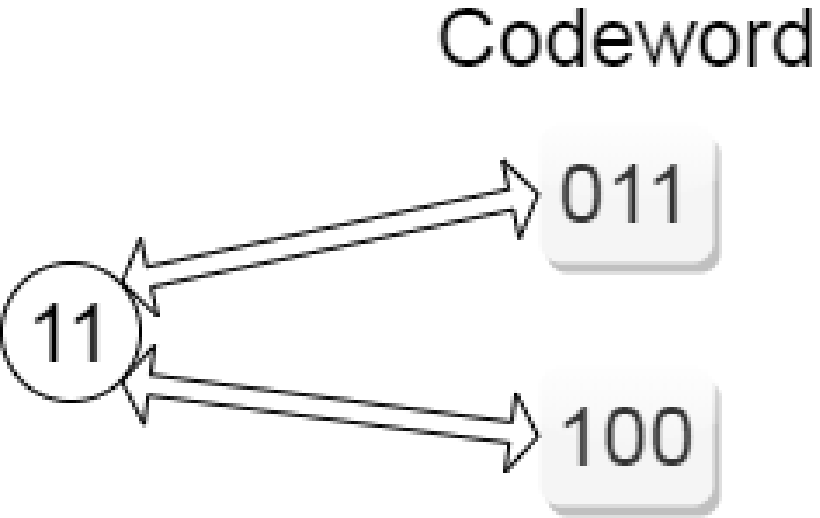}}
\subfigure[\sysname{} writes public data only once but chooses the codeword used based on the bits of the data to hide. This enables it to sureptitiously hide information even in the presence of a powerful multi-snapshot adversary.]{\label{fig:sub3}\includegraphics[width=.95\linewidth]{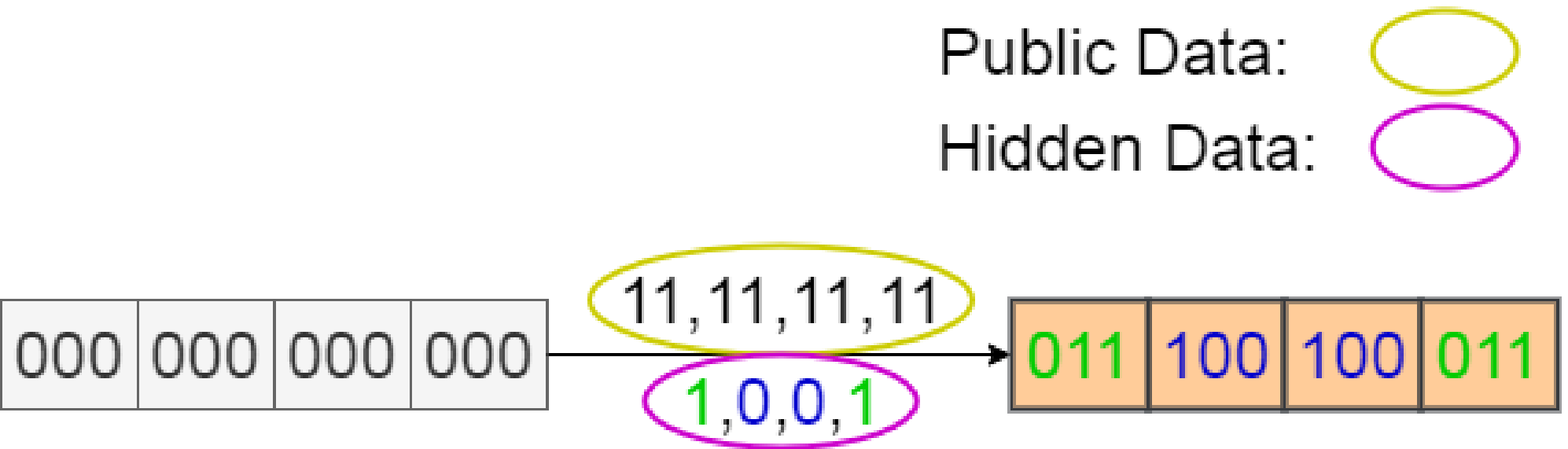}}
\caption{\small (a) Writing data multiple times using a simple WOM code. (b) WOM codes allow multiple codewords for the same data. (c) \sysname{} hides data by deciding the written codewords based on the data bits to be hidden. The resulting final physical state ($011,100,100,011$) is identical to the physical bits in Figure \ref{fig:sub1} resulting from two innocent writes.
% A WOM code (Table \ref{tab:wom_eg}) allows the same physical bits to be written to multiple (e.g., two) times. In Figure \ref{fig:sub1} two consecutive writes of different data ($10,00,11,01$ followed by $11,11,11,11$) can go forward in the same underlying physical locations. The end physical state is $011,100,100,011$. This is possible because (Figure \ref{fig:sub2}) the WOM code allows data $11$ to be represented by either $100$ and $011$ underlying bit configuration (``codeword''). \sysname{} writes public data once but chooses the codeword used based on the bits of the data to hide. This enables it to sureptitiously hide information even in the presence of a powerful multi-snapshot adversary. In the example in Figure \ref{fig:sub3}, the resulting final physical state hiding bits $1,0,0,1$ is the same ($011,100,100,011$) as the physical bits in Figure \ref{fig:sub1} resulting from two innocent writes. An adversary observing this final physical state cannot tell whether it is the result of two innocent sequential public writes as in \ref{fig:sub1} or of an information hiding operation as in \ref{fig:sub3}.
}
\label{fig:idea}
\vspace{-0.4cm}    
\end{figure}

The key idea in \sysname{} is to store both public data and hidden data {\em in the same physical locations} using a special data encoding scheme that renders a sequence of bits encoding public + hidden data bits indistinguishable from a sequence of bits only encoding public data. 
Before detailing the data encoding scheme used in PEARL, we provide an example to demonstrate how WOM codes can be used to indistinguishably encode hidden data (Figure \ref{fig:idea}).

% a sequence of bits encoding public + hidden data bits indistinguishable from a sequence of bits only encoding public data. In this section, we show that it is indeed possible to design such a data encoding scheme by leveraging a special write-one-memory (WOM) code. 
% Before detailing the data encoding scheme in PEARL, we provide an example to demonstrate how WOM codes can be used to indistinguishably encode hidden data (Figure \ref{fig:idea}).

% we describe next a data encoding scheme leveraging certain write-one-memory code (WOM code) that makes this possible. 

WOM codes (details in Section \ref{sec:wom}) are special data encoding schemes that allow multiple writes to the same locations of write-once memory by writing to some of the yet-unwritten-to bits (from 0 to 1). 

A WOM code (Table \ref{tab:wom_eg}) allows the same physical bits to be written to multiple (e.g., two) times. In Figure \ref{fig:sub1} two consecutive writes of different data ($10,00,11,01$ followed by $11,11,11,11$) can go forward in the same underlying physical locations. The end physical state is $011,100,100,011$. This is possible because (Figure \ref{fig:sub2}) the WOM code allows data $11$ to be represented by either $100$ and $011$ underlying bit configuration (``codeword''). In the context of flash devices, WOM codes are used to increase the amount of data you can write to a block before it is erased.

Our newly proposed data encoding scheme in \sysname{} writes public data once but chooses the codeword used based on the bits of the data to hide. This enables it to sureptitiously hide information even in the presence of a powerful multi-snapshot adversary. 

For example, as illustrated in Figure \ref{fig:sub3}, since the first hidden bit is ``1'', ``011'' is written to the underlying physical cells for the first two public data bits ``11''. On the other hand, the second hidden bit is ``0'', and in this case ``100'' is written for the second two public data bits ``11'', etc. 
The resulting final physical state hiding bits $1,0,0,1$ is the same ($011,100,100,011$) as the physical bits in Figure \ref{fig:sub1} resulting from two innocent writes. An adversary observing this final physical state cannot tell whether it is the result of two innocent sequential public writes as in \ref{fig:sub1} or of an information hiding operation as in \ref{fig:sub3}. In other words, it simply cannot distinguish the two cases with any non-negligible advantage and thus determine whether any hidden data exists.

It is clear from the example above that WOM codes have certain desirable properties that could provide opportunities for a data encoding scheme hiding hidden bits in a public cover. However, designing a general purpose data encoding scheme based on WOM codes that enables data hiding and is secure against a powerful adversary is not trivial. Specifically, as we discuss later, (e.g., because of device state biases interfering with indistinguishability and more), not all WOM codes can be used to build suitable data encoding schemes {\em and} not all data encoding schemes derived from WOM codes can be re-purposed for data hiding securely. Therefore, we first need identify what types of WOM codes can be re-purposed for our goals and then build a data encoding scheme accordingly. 

% In addition, several practical issues also need to be addressed e.g., management of obsolete data, garbage collection etc.  

% Although the above example seems hiding the hidden data bits in a public cover, there are still a number of challenges designing the such a data encoding scheme to write hidden data secretly. 
% For example, is arbitrary WOM code can be converted into a data encoding scheme? Is there any requirement for the obsolete data which is overwritten in Figure \ref{fig:sub1}? Can any data encoding scheme converted from a WOM code securely hide the existence of hidden data?
% In the rest of this section, we discuss how to design a data encoding scheme which can write hidden data together with public data and hide its existence as well exploiting a WOM code.

We start with an introduction of WOM codes in Section \ref{sec:wom}. Then we demonstrate the feature of a special group of WOM codes -- WOM codes supporting a 1st partition -- that can be used to encode hidden bits within public messages in Section \ref{sec:wom-partition}. After that, we propose our strategy to convert a WOM code to a hidden data encoding scheme in Section \ref{sec:encoding-scheme}. Finally, we show that not all hidden data encoding scheme based on WOM codes can ensure the deniability of the existence of hidden data, and propose a WOM code that can be indeed re-purposed for PD in the presence of a powerful adversary.

\subsection{Write-Once Memory Code}   % should the wom code be here or next section
\label{sec:wom}

Write-once memory (WOM) was first introduced in 1982 by Rivest \etal \cite{rivest1982reuse} and models a storage medium consisting of (binary) cells which can transition from a ``zero'' state to a ``one'' state only once. WOMs are written to using WOM codes, I/O schemes designed for this invariant. 
The WOM model was then generalized \cite{fu1999capacity,fiat1984generalized} for storage media cells with more than two possible states. 
Further, ``t-write WOM codes'' are WOM codes that can write additional information into the same group of WOM cells multiple ($t$) times. The number of bits that can be written on the each write does not need to be the same.

For simplicity and without loss of generality, WOM codes with two states only are used in the rest of the paper. Further, for consistency, initial states of NAND flash cells are considered to be ``zero'' even if in many chips, empty NAND flash pages physically contains all bits of 1.

% The primary goal of an efficient WOM code is to maximize the information that can be stored in a given number of storage cells, e.g., composing a page or block.

It is important to note that physically, NAND flash memory features the WOM invariant. Indeed once a flash page is written, its unwritten cells (only) can accept a second write cycle. Several studies propose WOM codes for lifetime extension by reduction in SSD block erasures \cite{wom-fast,Berman2013RetiredpageUI,jacobvitz2012writing,jagmohan2010write}.

\begin{table}[!bh]
\begin{center}
\begin{small}
% \vspace{-0.2cm}
\begin{tabular}{ | c | c | c |}
\hline
 Data bits & 1st write & 2nd write \\
\hline
00 & 000 & 111 \\
\hline
01 & 001 & 110 \\
\hline
10 & 010 & 101 \\
\hline
11 & 100 & 011 \\
\hline
\end{tabular}
\caption{{\small A WOM code that allows 2 writes of 2 bits within 3 bits.
\label{tab:wom_eg}}}
\vspace{8.4pt}
\vspace{-0.9cm}
\end{small}
\end{center}
\end{table}

Table \ref{tab:wom_eg} shows a WOM code example that allows twice the encoding of different configurations of 2 information bits using only 3 physical storage cells/bits. As per the WOM invariant, the 3 physical cells only change from 0 to 1 in both writes (the initial bits are considered to be $000$ before any write). For example, if the 2 bit message that needs to be written in the first cycle is $10$, $010$ will be physically written to the 3 storage cells. A subsequent 2 bit message $01$ will result in a second physical write of $110$. As can be seen, this requires a single change: the first physical cell needs to be set as 1 in the second write ($010 \to 110$). This elegantly enables in-place updates. Changing $10$ into $01$ would not have been possible in NAND flash without an expensive ERASE operation which significantly reduces lifetime and increases latencies.

Note that at first glance, it may seem that all the 3 physical cells/bits would change when the 2 bit message remains the same (e.g. $01$) for both 1st and 2nd write. This is actually not the case. Since both $001$ and $110$ represent message $01$ after the 2nd write, no physical bits need to be set.

Further note that the physical bits written in the second write are context dependent. They relate not only to the message itself but also to the existing data in the written cells. This mandates a read before the second write to perform the encoding correctly. Fortunately, NAND flash reads are much faster than ERASE operations.

\subsection{WOM  code  supporting  a  1st  partition}
\label{sec:wom-partition}

\paragraph{Notations}
``t-write WOM codes'' are WOM codes that can write additional information into the same group of WOM cells multiple ($t$) times (named {\it ``1st write, 2nd write'', ...} ) before requiring an ERASE. Each write requires a read of the existing physical state context, a proper encoding of the new logical data (``message'') using this context, and finally a physical write of the encoded result. The logical message encoded in the 1st write is called {\it ``1st message''} and the encoded result is called ``\firstrecord'', and so forth. For the sake of simplicity, and w.l.o.g. we consider only 2-write WOM code which has the same message space in both writes in the rest of the paper.

% \chen{The 2-write WOM code we considered maps message $m \in \Set{0,1}^k$ to become certain \womrecord{} $c_i$ in either write. }

Let $c_i \in C$ -- where $C = \Set{0,1}^n$ and $1 \leq i \leq 2^n$ -- denote the \womrecords{} of a n-bit WOM code. 
For example, for the WOM code example in Table \ref{tab:wom_eg}, we have $C= \lbrace c_1=000, c_2=001, c_3=010, c_4=011, c_5=100, c_6=101, c_7=110, c_8=111 \rbrace $.

For any two elements $c_x, c_y \in C$, the relationship $c_x \unrhd c_y$ is defined by the condition that $c_x[i] \ge c_y[i]$ for all $i \in [1, n]$, where $c[i]$ is the $i$-th bit of $c$. This is related to the fact that an unset flash bit can be easily set without requiring a page ERASE but not vice-versa. 
Then, a general definition for a 2-write WOM code can be given as follows \cite{yaakobi2012codes}: 

\begin{definition}[2-Write WOM Code]
\label{def:wom1}
A $(k,n)$ 2-write WOM code, denoted as $(k,n)$-$\mathsf{WOM}_2$, is an encoding scheme with message space $\Set{0,1}^k$ and codeword space $\Set{0,1}^n$ consisting of four algorithms $(\enc_1, \enc_2, \dec_1, \dec_2)$ that satisfy the following properties:
\begin{enumerate}
\item $\enc_1$: $\Set{0,1}^k \to \Set{0,1}^n$
\item $\enc_2$: $\Set{0,1}^k \times \Set{0,1}^n \to \{0, 1 \}^n$, and $\enc_2(m,c) \unrhd c$ for all $(m, c)$ 
%\in \Set{0,1}^k \times \Set{0,1}^n$.
\item $\dec_1$: $\Set{0, 1}^n \to \Set{0,1}^k$, and $\dec_1(\enc_1(m))=m$ for all $m$
%\in \Set{0,1}^k$ 
\item $\dec_2$: $\Set{0, 1}^n \to \Set{0,1}^k$, and $\dec_2(\enc_2(m,c))=m$ for all $(m, c)$
\end{enumerate}
\end{definition}

Informally, for the 1st write, any message is associated with a unique \womrecord{}. The 2nd write is a bit more tricky since the \secondrecord~ to be written depends not only on the {\it 2nd message} but also on the existing data (\firstrecord~) present in that location. Different values may end up being written i.e., $\enc_2(m, c_i)$ may be different from $\enc_2(m, c_j)$ for $c_i \neq c_j$. As a result, one message could be represented by more than one possible \womrecord{} after the 2nd write. For example, wrt. Table \ref{tab:wom_eg}, the message $00$ is always written as $000$ in the 1st write, but in the 2nd write it may be represented as either $000$, if the 1st written message was $00$, or $111$ otherwise.

$\enc_2()$ may have many different forms. We discovered multiple WOM codes that can represent a message using multiple \womrecords{} in the 2nd write. Our insight then is to use this degree of freedom in the choice of the \womrecord{} in the 2nd write to encode hidden information sureptitiously. For example, {\em two} \womrecord{} choices enable the encoding of {\em one} hidden bit. Generally, a choice of $2^m$ 
\womrecords{} allow the encoding of $m$ hidden bits. A simple encoding convention would be that using the $i$-th \womrecord{} choice indicates an encoded hidden value of $i$. 

In the rest of this paper, for simplicity, and w.l.o.g. we consider WOM codes with two choices only, i.e., which can encode {\em one} hidden bit through the encoding of each $k$ bit (public) message. These WOM codes have the following properties: (i) each message can be mapped to 2 \womrecords{} in the 2nd write, and 
% (ii) given arbitrary \firstrecord{},  
(ii) each codeword is corresponding to a few \firstrecords{}.
% \todo{what does this second point mean? }.
We call these WOM codes {\it WOM Codes Supporting A 1st Partition}:

\begin{definition}[WOM Code Supporting A 1st Partition]
\label{def:wom2}
Let $C_1$ denote the set of all \firstrecords{} for all possible messages, i.e. $C_1 = \{ \enc_1(m)\}_{m \in \Set{0,1}^k}$. Consider also a partitioning function $\prt(m)=(A_m, B_m)$ which on input $m \in \Set{0,1}^k$, outputs two sets $A_m$ and $B_m$, forming a partition of $C_1$, namely $A_m \cap B_m = \emptyset$ and $A_m \cup B_m = C_1$.

%The size of the two output sets are the same ($|A_m| = \|B_m|$).

Then, a $(k,n)$-$\mathsf{WOM}_2$ code $(\enc_1, \enc_2, \dec_1, \dec_2)$ is said to ``support a {\em 1st partition}'' if:
\begin{equation}
    \enc_2(m, c)= 
\begin{cases}
    \w_a(m), & \text{if } c \in A_m\\
    \w_b(m), & \text{if } c \in B_m
\end{cases}
\label{eq:1}
\end{equation}
where $A_m$ and $B_m$ are the 1st and 2nd output of $\prt(m)$, respectively and $\w_a(m)$ and $\w_b(m)$ are valid WOM code-specific functions that map input messages $m \in \Set{0,1}^k$ to \womrecords{} in $\Set{0,1}^n$.
\end{definition}

% \todo{You need to provide an intuition of what happens here in 1-2 paragraphs. What does this achieve? Why is it defined like this? don't forget to mark the new text blue.}

% \todo{also $\w_a(m)$ and $\w_b(m)$ are undefined basically so this is very vague and $enc_2$ ends up being undefined also as a result. overall i am still missing what is happening here. I added some text to the definition but i am unsure if it is clearer}

Note that for a valid 2-write WOM code -- which requires $\enc_2(m ,c) \unrhd c$ -- we must have $\w_a(m) \unrhd c_a$ for any $c_a \in A_m$ and $\w_b(m) \unrhd c_b$ for any $c_b \in B_m$. 

Specifically, we call a WOM code supporting a 1st partition where $|A_m| = |B_m|$ for all $m \in \Set{0,1}^k$, a {\em WOM code supporting an equal partition}. 

Table \ref{tab:wom_eg} illustrates a WOM code supporting a 1st partition, but not an equal partition. Consider $C_1=\Set{c_1=000, c_2=001, c_3=010, c_4=100}$. For each message $m$, $A_m = \Set{\enc_1(m)}$, and $B_m = C_1 \setminus A_m$, $w_a(m)$ and $w_b(m)$ are the \womrecords{} in the row corresponding to message $m$ where $w_a(m)$ is in the second column and $w_b(m)$ is in the third column. It can be seeen that $|A_m|=1$ and $|B_m|=3$ for any message $m$. Thus, the WOM code does not support an equal partition.

As we will see later, the ability to support an equal partition enables the design of a plausible deniability mechanism in which the resulting distribution of the written bits does not leak information about the encoded hidden data. 

\subsection{Hidden data encoding scheme}
\label{sec:encoding-scheme}

\paragraph{Hidden data encoding}
As discussed above, for a WOM code supporting a 1st partition, encoding a ``hidden'' data bit $h$ within a $k$-bit ``public'' message $p$ can be achieved by using the bit to decide on the choice of \womrecord{} to write in the 2nd write. 

Then, more generally, the written data $\enc(p, h)$ is a function of both the public message $p$ and the hidden message $h$. Further as we will see, there exists a relationship between the existing data $c$ and the resulting encoding. 

% We call them ``public message $p$'' and ``hidden message $h$'' in the rest of the paper for consistency. The same group of {\it WOM codes} which are used in the {\it 2nd write} can be take advantage of to encode the public and hidden messages. The public message behaves like the message $m$ while the hidden message takes the role of the code $c$. 
%The public message decides the subset of {\it encoded data} that the final result can be chosen from, while the hidden message decides the final code.
We call the encoded result the {\it ``\fullcode''} and the write of the \fullcode~ is called a {\it ``full write''} to distinguish it from a 2nd write of a public-only message.
% for distinction\todo{distinction from what?}. 
Then, the corresponding simplified encoding function is: 
\begin{equation}
    \enc(p, h, c)= 
\begin{cases}
    w_a(p), & \text{if } h = 0\\
    w_b(p), & \text{if } h = 1 
\end{cases}
\label{eq:2}
\end{equation}

As mentioned earlier, for simplicity, and w.l.o.g. we consider WOM codes with two \secondrecords{} choices only (as in equation \ref{eq:1}), i.e., which can encode {\em one} single hidden bit with each $k$ bit (public) message. 

Unfortunately there are no free lunches and, as will be detailed later, the \fullcode~ can only be written to empty pages with all 0s. In other words, one page cannot be written-to twice anymore. Effectively, the hidden bit is encoded at the cost of the ability of the WOM code to accept additional information before the next ERASE.
% \todo{and where are you explaining the expense? the next sentence doesn't seem to talk about any overhead/expense etc}

Note that one of $w_a(p)$ and $w_b(p)$ are written regardless of the existing data $c$. And, as discussed above, for a valid 2-write WOM code -- which requires $\enc_2(m ,c) \unrhd c$ -- we must have $\w_a(p) \unrhd c$ and $\w_b(p) \unrhd c$ for all possible $p$.
%
% The above encoding scheme requires that the both the {\it WOM code} $w_a(p)$ and $w_a(p)$ should be able to be written without violating the constraints of the directed state changing, since the hidden message can be either 0 or 1.
%Thus, we skip the {\it 1st write} to ensure the existing data contains only bit 0. In other words, only an empty physical page can be written with the above hidden data encoding scheme and the page can only be written one time before ERASE.
%
This is only possible if the existing data $c$ is all 0s, i.e., the encoding works only on empty physical pages, and pages can only be written once before requiring an ERASE.

\begin{table}[!t]
\begin{center}
\begin{small}
%\vspace{-0.3cm}
\begin{tabular}{ | c | c | c | c |}
\hline
 \multirow{2}{*}{Data bits} & \multirow{2}{*}{1st . write} & \multicolumn{2}{c|}{2nd write} \\
\cline{3-4}
  &  & Hidden 0 & Hidden 1 \\
\hline
00 & 000 & 000 & 111 \\
\hline
01 & 001 & 001 & 110 \\
\hline
10 & 010 & 010 & 101 \\
\hline
11 & 100 & 100 & 011 \\
\hline
\end{tabular}
\caption{{\small A WOM code that provides a trade-off between wear leveling and plausibly deniable information hiding. Within a 3 cell area, between ERASEs, allow either: (i) {\em two} writes of 2 bits each, or (ii) {\em one} write of a 2 bit public message plus a 1 bit hidden message.
\label{tab:wom_eg-pd}}}
\end{small}
\end{center}
\end{table}

Table \ref{tab:wom_eg-pd} adds the hidden bit encoding cases to the WOM code in Table \ref{tab:wom_eg}.
% Partition sets $A_m$ and $B_m$ are not explicitly listed since it is easy to learn which \womrecord{} can be written given the existing data.
Between ERASEs, 3 flash cells can be written to either: (i) twice with 2-bit {\it public  messages} or once with a 2-bit {\it public message} and a 1-bit {\em hidden message}. 

The first case corresponds to public-only operation in which the same set of cells can be reused for a 2nd write between ERASEs and the second case corresponds to the case of hidden operation in which hidden messages are to be encoded plausibly deniable.

\begin{figure}[t!] 
\centering
\includegraphics[width=6cm]{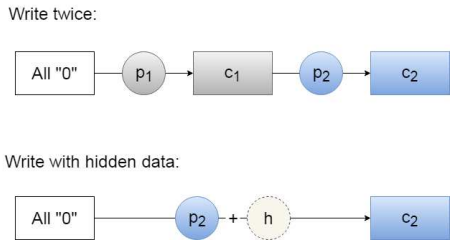}
\caption{\small A empty page can be written to twice in two sequential public writes $p_1 \& p_2$, or once with an encoding combining one public message $p$ and one hidden message $h$. The resulting state is the same $c_2$.}
\label{fig:indistinguish}
\vspace{-0.4cm}    
\end{figure}

The first case corresponds to a sequence of public writes, e.g., an initial write and subsequent updates to the same location. Starting with an empty page (e.g., of 3 bits for simplicity), a public message $p_1$ can be written into an empty page firstly (as $c_1$) in a {\it 1st write} (Figure \ref{fig:indistinguish}). The encoding used is defined by columns 1\&2 in Table \ref{tab:wom_eg-pd}. A subsequent public message $p_2$ can be written to the same page (as $c_2$) in a {\it 2nd write}. Columns 1,3 and 4 in Table \ref{tab:wom_eg-pd} determine the final written state, as a function of \firstrecord~ $c_1$. Note that after the {\it 2nd write} the first public message $p_1$ will not be available any more. Only $p_2$ can be decoded from $c_2$. 

In the second hidden operation case, both public ($p_2$) and hidden ($h$) messages determine the encoding that gets written ($c_2$) in a {\it full write}. $c_2$ is determined by columns 1,3, and 4 of Table \ref{tab:wom_eg-pd}. Once written, both $p_2$ and $h$ can be decoded from $c_2$.

\subsection{WOM coding \& PD}
\label{sec:skew}

\paragraph{0-1 Distribution Skew}
% With the encoding scheme explained above, we can take advantage of a WOM code supporting 1st partition and finally encode one hidden bit with two normal bits into 3 storage bits, at the expense that the 3 storage bits can only be written once. However, does this mean that we now have a secure scheme to hide the modification to the storage device due to the hidden data from a PD adversary? Unfortunately, the answer is NO.
%
While a step in the right direction, the proposed hidden data encoding results in a bias in the distribution of 1s and 0s in the written data when compared to a public-only operation. This can then be used e.g., by a multi-snapshot adversary to distinguish devices that contain hidden information from devices that do not.

% is not secure for the reason that the resulting \fullcode~ stored on the device will have a biased 0-1 distribution, compared with the original \secondrecord~. This may result into the deniability compromise when adversaries do statistic analysis. 

\begin{figure}[t!] 
\centering
\includegraphics[width=5.5cm]{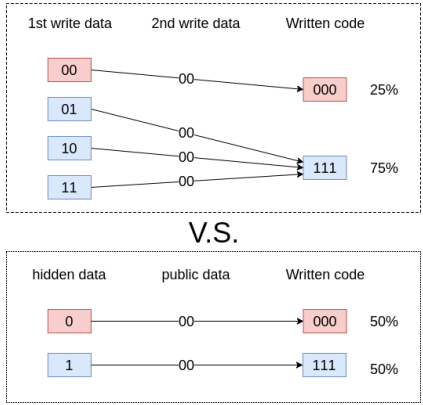}
\vspace{-0.1cm} 
\caption{{\small The WOM code in Table \ref{tab:wom_eg-pd} features a bias in the distribution of written 1s and 0s. The top part illustrates the resulting skew (0s:25\%, 1s:75\%) for public-only operations after two writes, whereas the bottom part illustrates the hidden operation (0s:50\%, 1s:50\%).}}
\label{fig:skew}
\vspace{-0.2cm}    
\end{figure}

Figure \ref{fig:skew} shows a simple example of this bias for the example WOM code in Table \ref{tab:wom_eg-pd}. 
Consider a 2 bits public message $00$. In public-only operation mode, the message is written in a {\it 2nd write}, and the \secondrecord~ will be $000$ if the {\it 1st message} residing there was $00$, or $111$ if {\it 1st message} there is either $01$, $10$, or $11$. If the {\it 1st message} written is overall uniformly distributed, the ratio between the occurrence of $000$ to $111$ in the public-only operation storage device should be 1:3. 

However, in the case of a hidden operation, the \fullcode~ ends up being $000$ for a hidden bit of $0$, or $111$ for a hidden bit of $1$. Thus, for an overall uniformly distributed hidden message, the ratio between the occurrence of $000$ to $111$ in the storage device will be 1:1. 

Given this bias, an adversary can do statistic analysis based on the public data on the storage data and
observe a difference between the expected and observed distribution of 0s and 1s.

A counter-argument to be made is that the public operation mode was considering the case of two writes, and in practice numerous pages may end up being written only once. This may be true, however, given the existence and benefits of the WOM encoding in the system, it is reasonable to expect that in many cases, the device converges to a state where most cells have been overwritten at least once. 
Also, while it is true one can plausibly claim the bias was inherent in the data itself, the security argument is weakened overall. 

% \vspace{-0.1cm} 

\paragraph{WOM Code Supporting an Equal Partition}
Thus, the question inevitably arises: can we do better? How can we overcome this bias?
%
% Although WOM codes supporting 1st partition cannot ensure indistinguishability between a full write with two consecutive WOM writes to a page, 
%
{\em The answer is WOM codes supporting an equal partition.}
%
% One lesson we can learn from the above analysis is that not all WOM codes can be used to write hidden bits without arousing any suspicion. The hidden data encoding scheme would be secure against a PD adversary only if the WOM code is ``matched'' with the hidden data in certain way. Here we explain what is the feature a WOM code should have in order to make it possible to extend it to be a secure hidden bit encoding scheme. 

To see why that is the case, consider that the bias comes directly from the difference in the probability distribution of \secondrecord~ and \fullcode~.
Reusing the same group of \womrecords{} in the {\it 2nd write} for the hidden data encoding ensures that the \secondrecords~ and \fullcodes~ are indistinguishable by inspecting the individual codes. It is an overall probability distribution that may give the existence of hidden data away.

Note that the probability distribution of \secondrecord~ depends on the number of elements in sets A and B (see equations \ref{eq:1} and \ref{eq:2}), while the probability distribution of \fullcode~ is decided by the distribution of hidden bit $h$. And, since hidden data in a PD system is highly likely to be encrypted, $h$ ends up uniformly distributed. 

To eliminate the bias, sets A and B need to contain the same number of elements for any arbitrary messages. In other words, the WOM code needs to support an equal partition.

\vspace{-0.1cm}
\begin{lemma}[]
\label{lem0}
The hidden data encoding scheme based on a WOM code supporting an equal partition ensures that an adversary cannot distinguish the  \secondrecord~ that encodes public message $p$ from the \fullcode~ that encodes both public message $p$ and hidden message $h$.
\end{lemma}
\vspace{-0.2cm}

% The intuitive method to encode the hidden bits is to indicate the hidden bit with the choice of written code. This requires that the occurrence rate of the written codes corresponding to hidden bit 0 is the same as the occurrence rate of the written codes corresponding to hidden bit 1. For a 2-write WOM code that can represent n messages ($logn$ bits) in the second write, suppose that message $k (1 \leq k \leq n)$ can be written as $m_k$ codes, it requires that $\forall k (1 \leq k \leq n)$, there exists a partition of the $m_k$ codes such that the written code in each partition can be result from half of the 1st write data. 

\begin{table}[!t]
\begin{center}

\begin{small}
% \vspace{-0.3cm}
\begin{tabular}{ | c | c | c | c | c |}
\hline
\multirow{2}{*}{ } & \multirow{2}{*}{Data bits} & \multirow{2}{*}{1st write} & \multicolumn{2}{c|}{2nd write} \\
\cline{4-5}
 &  &  & Hidden 0 & Hidden 1 \\
\hline
 0 & 000 & 00000 & 11110 & 10011  \\
\hline
 1 & 001 & 00001 & 11001 & 10110 \\
\hline
 2 & 010 & 00010 & 11010 & 10101 \\
\hline
 3 & 011 & 00100 & 11100 & 01111 \\
\hline
 4 & 100 & 01000 & 11111 & 01101 \\
\hline
 5 & 101 & 10000 & 11101 & 01110 \\
\hline
 6 & 110 & 11000 & 11000 & 10111 \\
\hline
 7 & 111 & 10100 & 11011 & 10100 \\
\hline
\end{tabular}
\caption{\small{$(3,5)$ WOM code supporting an equal partition allowing, within 5 bits: two subsequent public writes of 3 bits, or one write of 1 hidden bit and 3 public bits.
\label{tab:wom_pd}}}
%\vspace{8.4pt}
\vspace{-0.8cm}
\end{small}
\end{center}
\end{table}

% * Data                 First                           Second
%  * 0. 000               00000           11110(367)2             10011(1)045
%  * 1. 001               00001           11001(146)0             10110(237)5
%  * 2. 010               00010           11010(246)0             10101(137)5
%  * 3. 011               00100           11100(567)0             01111(12)34
%  * 4. 100               01000           11111(2567)             01101()0134
%  * 5. 101               10000           11101(1567)             01110(2)034
%  * 6. 110               11000           11000(46)05             10111(1237)
%  * 7. 111               10100           11011(1246)             10100(37)05

%\paragraph{Secure WOM code for hidden data encoding}
Table \ref{tab:wom_pd} defines a $(3,5)$ WOM code supporting an equal partition. For public operation, it allows writing 3 bits of data twice to 5 storage cells. In hidden operation, 1 hidden bit and 3 public bits can be encoded together in 5 storage cells. 

% One invariant for this code is that each of the \secondrecords{} in the ``2nd write'' column (sub-columns 4 and 5) corresponds to four {\it 1st messages}. 
One invariant for this code is that for each \secondrecord{} $c_2$ in the ``2nd write'' column (sub-columns 3 and 4), there exist four \firstrecords{} $c_1$ for which $c_2 \unrhd c_1$, i.e., that can be overwritten to get to $c_2$. In other words, the size of the sets A or B in this WOM code is 4.

For example, considering $w_a=11110$ and $w_b=10011$ -- both of which can be used to represent public message $m=000$ in a {\it 2nd write} during public operations -- the corresponding set $A$ is $A_{000}=\Set{00100, 01000, 11000, 10100}$, composed of the \firstrecords{} for messages $\Set{3,4,6,7}$. Similarly, set $B$ is $B_{000}=\Set{00000, 00001, 00010, 10000}$ composed of the \firstrecords{} for messages $\Set{0,1,2,5}$.
% we have as follows: $11110$ corresponds to (can be written after\todo{??? is ``corresponds to'' defined somewhere?}) {\it 1st messages} $\Set{2,3,6,7}$, while $10011$ corresponds to {\it 1st messages} $\Set{0,1,4,5}$. 
% These sets are of the same sizes, and 
As a result, both $11110$ and $10011$ are equally likely to appear in the written device state -- for either public and/or hidden operation modes. 
% \todo{please add one column at beginning with 0,1,2 ... 7 so we know what you are refering to}

Based on the WOM code in Table \ref{tab:wom_pd}, we design \sysname~, a plausibly deniable FTL that securely processes the I/O request from the upper layers, manages the unavoidable inherent mappings from logical to physical pages and reclaims pages occupied by the obsolete data. More importantly, \sysname{} ensures that adversaries cannot detect the existence of hidden data by probing multiple device snapshots.

\section{Security Requirements for \sysname{}}
\label{sec:requirement}

% The property of WOM code presented in Section 5 serves as the main idea behind our PD scheme. However, turning it into a workable PD solution requires extra work. In this section, we highlight the related issues and present corresponding solutions. 
The hidden data encoding scheme presented in Section \ref{sec:pdwom} ensures that physical pages containing both public and hidden data are indistinguishable from pages containing public data written as \secondrecords~.
However, turning it into a workable PD solution that can protect hidden data from the coercive adversary described in Section \ref{sec:model} requires extra work.
% Yet this is not enough to ensure secure PD.  
Specifically, a multi-snapshot adversary can observe not only the state of individual pages, but also state changes across multiple snapshots. Generally speaking, over time, an adversary can learn (1) what kind, and (2) where page state transitions happen. A plausibly deniable FTL needs to ensure that this information does not leak the existence of hidden data.

This is made even more difficult by internal characteristics of NAND flash for which page state transitions are not independent from each other. For example, data updates are performed via an out-place scheme rather than an in-place scheme (updated data is written to a new location rather than where the old data resides). As a result, pages where the up-to-date data is written becomes valid while at the same time the page where the outdated data resided becomes invalid.

% For example, up-to-date versions of logical page data likely end up being written after \todo{you mean at a higher offset on the disk?} the locations containing the outdated version of the same logical page. 
% As a result, a plausibly deniable FTL needs to guarantee that all the page states as a whole does not leak the existence of hidden data. % you already say that above 

To mitigate this, we first explore the page states and the page state transitions in the case of deploying the WOM code. We then introduce key requirements for a secure plausibly deniable FTL. Finally, we provide an efficient solution. The idea is to smartly ``cloak'' hidden data within plausible public data so that the hidden data induced page state transitions can be plausibly explained as a result of public requests.

\paragraph{Page States}
NAND flash contains three types of pages: empty, valid, and invalid.
A ``valid'' page contains active data, whereas an ``invalid'' page's data is obsolete and can be erased. 

In the case of a 2-write WOM code, NAND flash pages can be categorized at a finer granularity. 
% \chen{todo}
Firstly, based on their current encoding,  pages can be categorized as either ``1st write'' or ``2nd write'' pages. 1st write pages contain only public data while 2nd write pages may contain both public data and hidden data. Note that in this case, a page storing both public data and hidden data is still called a 2nd write page although the page is literally written only once.

Secondly, a page can be either valid or invalid depending on the status of the data stored inside. However, since the public data and hidden data in the same page may have different status we need to further distinguish things.  We use ``up-to-date'' and ``out-of-date'' to indicate data status. Then, since the existence of hidden data should not be exposed to adversaries, a page is denoted as valid as long as the public data there is up-to-date, regardless of whether any hidden data coexists or whether the hidden data is out-of-date. 
% Meanwhile, the public data and hidden data stored in the same physical page can has different status. For better distinction, we use ``up-to-date'' and ``out-of-date'' to indicate whether the data is valid or not. A page is denoted as a valid page as long as the public data there is still up-to-date, regardless of whether any hidden data coexists or whether the hidden data is out-of-date.
Note that a valid page may contain out-of-date hidden data while an invalid page may contain up-to-date hidden data. 

Thus, in summary, a page can be in any of the 5 states: empty, 1st write valid (V1), 1st write invalid (I1), 2nd write valid (V2), and 2nd write invalid (I2).
% \todo{this sentence doesn't make any sense -- for example 1st write valid cannot considered as empty can it ? -- also there are only 4 states listed here ...}
% 
% \chen{Moreover, based on the different reason why data becomes out-of-date, a V1 page can be further divided to be UI page and TI page}
% 
Each physical page transitions between the 5 states directly or indirectly. There may be more than one possible reason for a page to change from one state to another. For example, an empty page may be turned into a V2 page directly because of a {\it full write} (defined in Section \ref{sec:pdwom}), or it can become a V2 page indirectly by first being a V1 page, then an I1 page and finally a V2 page. Thus, the first requirement is shown as Requirement \ref{req1}.

\begin{requirement}[]
\label{req1}
The presence/absence of hidden data is never the only possible reason for a page state transition. Note that a hidden data encoding scheme based on a 2-write WOM code is designed to intrinsically guarantee this.
\end{requirement}

% Then, a {\it first requirement is that the design ensures hidden data is never the only possible reason for a page state transitions}. Note that the hidden data encoding scheme based on 2-write WOM code is designed to intrinsically guarantee this.

\begin{figure}[t!] 
\centering
\vspace{-0.4cm}
\includegraphics[width=4.8cm]{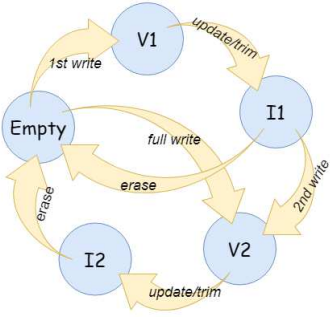}
\vspace{-0.6cm} 
\caption{\small Page state transition diagram using the 2-write WOM code. A page written twice with public data can transition through all 5 states while a page written once with public and hidden data skips states V1 and I1. It is also possible that a page is recycled right after it is written only once with public data (state I1 to Empty directly).}
\label{fig:state}
\vspace{-0.4cm}    
\end{figure}

Figure \ref{fig:state} depicts the page state transition graph for the above 5 states. Transitions are triggered by either the logical requests from the host or the built-in functions of the NAND flash (e.g., garbage collection).  Public data can be written to either an I1 page or an empty page, resulting in a V2 page or a V1 page, respectively 
% -- this corresponds to the state transition from $S_1$ to $S_2$ or from $S_3$ to $S_4$.
Hidden data can only be written to an empty page under a cloak of some public data. This is depicted as a full write that transitions a page from state empty to state V2 directly. 
Garbage collection brings all the pages in a target block back to state empty by an ERASE operation. Before erasing, the up-to-date data in those pages need to be relocated elsewhere, while the state of the page may transition from V1 to I1 or from V2 to I2. Other operations that render data out-of-date include logical data updates and the TRIM operation.

Figure \ref{fig:state} illustrates the fact that a page can transition freely from one state to another independently of the existence of hidden data -- all page state transitions can be plausibly explained by public data operations.

Moreover, the plausible public data that can be used as the ``cloak'' is not unique and in fact has quite a bit of entropy. For example, as Figure \ref{fig:indistinguish} depicts, 
writing hidden data $h$ + public data $p_2$ ends up being the same as writing public data $p_1$ + $p_2$. As the \firstrecord~ is completely overwritten by the \secondrecord~, a relatively large set of public data messages can be plausibly provided as a candidate for $p_1$. 
For a $(k,n)$ WOM code supporting an equal partition and pages contain $n \cdot x$ bits, there are  $2^{(k-1) \cdot x}$ possibilities for $p_1$.
% \todo{how many? what is the relationship between the size of this set and the code size? or some other meaningful relationship?} 

More specifically, as an example, consider a physical page of 10 physical bits. In \sysname{} the page can contain 6 bits of public data and 2 bits of hidden data. If the observed physical page data $1100010101$, then $p_2=110010$ and $h=01$ according to Table \ref{tab:wom_pd}.  An attacker obtaining a snapshot aiming to determine the value of public data $p_1$ can at most know is that the first 3 bits of $p_1$ are a value in set $\Set{000,100,101,110}$ (the messages corresponding to \womrecords{} in set $A_{110}$) and the last 3 bits of $p_1$ are a value in set $\Set{001,011,101,111}$ (the messages corresponding to \womrecords{} in set $B_{010}$). As a result, $p_1$ has 16 ($4^2$) possible values in total. In reality, a larger page with more physical bits (e.g. $5 \times 1000$ bits) results into many possibilities (e.g. $4^{1000}$). For further security, this can then be used to select the most semantically plausible values of $p_1$, e.g., by selecting marching terms from an English dictionary.

\paragraph{Page Operation Priority}
%%
% \chen{where to write and where to GC}
The WOM code based hidden data encoding scheme ensures multi-snapshot adversaries cannot tell whether hidden data exists or not by observing state transitions of any single physical page. However, by observing aggregated state transitions of a set of pages over time, it may still be possible for an adversary to detect the existence of hidden data according to where page state transitions happen (which page is written to and which page is erased). 

For example, if pages containing up-to-date hidden data have a lower ERASE priority during garbage collection compared to pages containing no hidden data, an adversary could tell whether the hidden data exists through the order in which physical pages get erased. 
% More important, with two snapshots of the device, an adversary may be able to infer the history of page state transitions happened between the two snapshots. This makes it harder to plausibly deny the existence of hidden data. 
Thus, a second requirement can be concluded as Requirement \ref{req2}.

\begin{requirement}[]
The priorities assigned to blocks according to which they are erased during garbage collection is not be related to the location, state or existence of hidden data.
%The erasing priorities of blocks 
\label{req2}
\end{requirement}

% Thus, {\it to ensure PD, a second requirement is that the erasing priorities of blocks should not be related to the location, state or existence of hidden data.} 

Moreover, as illustrated in Figure \ref{fig:indistinguish}, hidden data $h$ in a 2nd write page can be plausibly denied as a sequence of public operations: $p_1$ written to an empty page, and $p_2$ written to an I1 page. Solid reasons should exist to justify why $p_1$ is not written to any other I1 page and $p_2$ is not written to any other empty page etc.
This derives the Requirement \ref{req3}.
\begin{requirement}[]
The presence/absence of hidden data is never the only possible reason for the presence of any public data in a 2nd-write page.
\label{req3}
\end{requirement}

% Then, {\it a third requirement is that the FTL ensures hidden data is never the only possible reason for the ``cloak'' public data to reside in a 2nd write page.}

Fulfilling this requirement \ref{req3} efficiently is related to how writing priorities for empty pages and I1 pages are defined in an FTL (more details in Section \ref{sec:sys}).

Note that in order to maximize writing capacity and minimize wear/ERASE cycles, normally I1 pages usually have higher priority to be written to. Otherwise if empty pages are written first, then an I1 page may be erased before the 2nd write happens to it, which is a waste of writing capacity.

\section{\sysname~ Design}

% \subsection{\sysname~ Design}
\label{sec:sys}
\sysname{} is a FTL that satisfies all the security requirements introduced in Section \ref{sec:requirement}. It is designed based on DFTL (Section \ref{section:DFTL}). In this section, we firstly describe in detail the data structures used for logical-to-physical address translation and the page allocation mechanism. Based on them, we then introduce how \sysname{} deals with the public and hidden requests from the host and reclaims the obsoleted pages with garbage collection.

\vspace{-0.1cm}
\subsection{Address Translation}
% 
% \sysname~ manages both the public data and hidden data with page-level mapping.
\sysname{} manages the logical-to-physical mapping for public data and hidden data separately.
Similar to DFTL (Section \ref{section:DFTL}), two layer page-level maps are used.
The public data is managed by a public global translation directory (GTD) plus a few public translation pages, while the hidden data is mapped with a hidden GTD in addition with some hidden translation pages. 
The hidden GTD is stored in the SRAM together with the public GTD. If no power loss protection is built into the flash device, GTDs may get lost during sudden power loss, but can always be recovered by a full device scan. Furthermore, storing GTD on nonvolatile storage aids recovery from power-failure \cite{urgaonkar2008dftl}.

Translation pages are stored in the flash. 
% The public translation pages and the public data pages are regardes as public pages and  
Unlike in the case of DFTL, both translation pages and data pages are stored in the same group of blocks. The public translation pages are encoded as public data, while hidden translation pages are encoded as hidden data.

A cached mapping table (CMT) is used to cache the recently-used mapping information for both public data and hidden data. The corresponding public or hidden translation page will be updated in memory whenever any mapping entry is evicted from the CMT.

\begin{figure}[t!] 
\centering
\includegraphics[width=7.5cm]{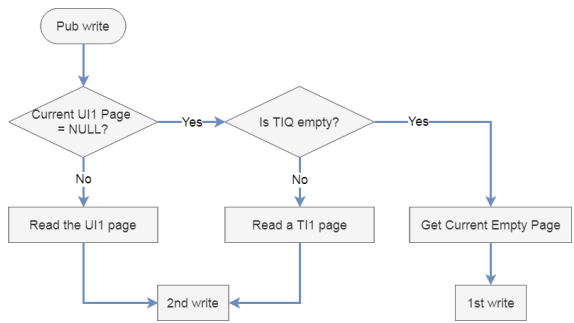}
\vspace{-0.2cm}  
\caption{\small The diagram about how \sysname{} allocates physical pages upon accepting a public write request. The priorities of physical pages are: 1st invalid pages in UIQ, 1st invalid page in TIQ, empty pages.}
\label{fig:alloc}
\vspace{-0.4cm}    
\end{figure}

\subsection{Page Allocation and Garbage Collection}
\label{sec:gc}
\begin{figure}[t!] 
\centering
\includegraphics[width=6cm]{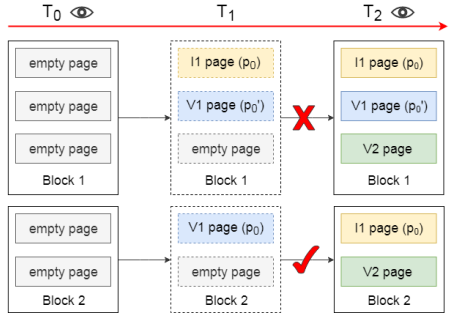}
\caption{\small Without hidden data, block 1 cannot plausibly transition from the state at time $T_0$ to the state at time $T_2$ because of the existence of the UI1 page -- the I1 page in block 1 is a UI1 page as its corresponding up-to-date data ${p_0}'$ is in the V1 page. In contrast, block 2 can plausibly transition between the states at times $T_0$ and $T_2$ regardless of the existence of hidden data, because an adversary cannot identify when a TI1 page becomes invalid -- the I1 page in block 2 is a TI1 page as it does not have corresponding up-to-date data in the block.}
\label{fig:example}
\vspace{-0.2cm}    
\end{figure}

\paragraph{Page Allocation}
\sysname~ uses three variables to track candidate pages for writing: a {\it Current Empty Page}, a {\it Current UI1 Page}, and a {\it TI1 page Queue (TIQ)}. UI1 pages are I1 pages caused by logical data updates. Whenever the public data in a V1 page gets updated, rather than updating in place, the up-to-date data is written to another page and the V1 page becomes a UI1 page. 
TI1 pages are I1 pages resulting from TRIM operations that delete data. The deleted data in a TI1 page does not have corresponding up-to-date data in any other pages of the device. 
We distinguish UI1 from TI1 pages since an adversary can infer when a UI1 page becomes invalid with only one device snapshot (detailed later and in Figure \ref{fig:example}). 
% TIQ is a queue that records the TI1 pages in the device.  % you already say that. 
Finally, a free block list ({\it FBL}) is used to track empty blocks in the device. 

In \sysname~ UI1 pages have the highest priority to be written to. As a result, since they always get written to first, there ends up being at most one UI1 page in the device, tracked by the {\it Current UI1 Page} record. Further, TI1 pages have a higher priority than empty pages to be written to. Overall, public data will be written to an empty page only if the {\it Current UI1 Page} is NULL and the TIQ is empty. Figure \ref{fig:alloc} illustrates the page allocation rule for public data.

In contrast, it should always be the {\it Current Empty Page} that is allocated for a hidden data write. This makes it possible for an adversary to infer whether a 2nd write page contains hidden data by inferring whether there exists any I1 page (either UI1 or TI1) when the 2nd write page is written to. Moreover, UI1 pages impose different threats compared to TI1 pages, which can be illustrated with Figure \ref{fig:example} as follow.

For a UI1 page that becomes invalid before any hidden data is written, an adversary would always know that it is invalid when the 2nd write page is written to. 
This is straightforward if the adversary can observe the UI1 page before writing the 2nd write page.
Besides, the example of block 1 in Figure \ref{fig:example} explains that this is also true even if the adversary can access the device only after the hidden data is written.

The upper half of Figure \ref{fig:example} lists snapshots of block 1 over time. The adversary observe the block at time $T_0$ and $T_2$. All three pages are empty at time $T_0$. And they are in state I1, V1 and V2, respectively at time $T_2$. The I1 page is a UI1 page as it contains the obsoleted data $p_0$ whose corresponding up-to-date data ${p_0}'$ is in the V1 page.
Based on the two snapshots, the adversary can infer that: 1) the first page must be a I1 page right after the second page is written to; 2) the third page should be still empty at that time (since pages in one block are written in order). Thus, the adversary can infer (although not directly observe) that there must exist an intermediate state where the there pages are in state I1, V1 and empty, respectively, which is depicted as the snapshot at time $T_1$. 
In this case, hidden data becomes the only possible reason for public data to reside in the V2 page rather than the I1 page in time $T_2$.

On the contrary, an adversary cannot tell whether a TI1 page becomes invalid before or after any hidden data is written as long as she cannot observe the TI1 page before the hidden write happens.
This can be demonstrated with
the example of block 2 in Figure \ref{fig:example}. 
Similarly, the adversary takes the snapshot at time $T_0$ and $T_2$. Then she observes the state transition empty $\to$ I1 in the first page, and the state transition empty $\to$ V2 in the second page.
% that the first page is changed from state empty to state I1 and the second page is changed from state empty to state V2.
The only intermediate state she can infer is that the first page was written to be a V1 page at certain time $T_1$. After that, the adversary has zero knowledge about whether the V1 page is invalidated first or the empty page is written first. Thus, it is completely possible for the second page to be the V2 page at time $T_2$ without any hidden data, as long as the second page is written before the first page becomes invalid.

Thus, to mitigate the possible leaks caused by I1 pages, two tweaks are used regarding the UI1 page and TI1 page: (i) before writing any hidden data, the {\it Current UI1 Page} is filled with public data; and (ii) all TI1 pages in the TIQ are written to (with public data either from user requests or from the block with the least number of valid pages) before on-event adversaries are allowed to take a snapshot. These prevent adversaries from detecting the hidden data by analyzing page allocation patterns.

%\blue{
\paragraph{Garbage Collection}
\sysname~ tags the {\em least active block}(s) (with the least number of valid pages) as the next victim block(s) for garbage collection. As detailed in Section \ref{sec:requirement}, the status of a page -- valid/invalid -- is independent of the existence and status of hidden data stored in that page. Thus, the selection of victim blocks does not leak any information to the adversary regarding the presence of hidden data.

Once a victim block is selected, \sysname~ first checks whether the {\it Current UI1 Page} is in the victim block. If yes, the {\it Current UI1 Page} is set to NULL. Then all the TI1 pages in the victim block are extracted from the TIQ. These two actions prevent data from being written to a block that will be erased soon. Then, up-to-date public and hidden data in the victim block are relocated to new pages using the same mechanism that is employed during write requests.
%
% The relocated data includes both public and hidden data in the victim block. 
% Hidden data can be encoded and written to the empty pages together with public data in the victim block or new public write request. Note that the public data that is paired with a hidden page could vary during garbage collection. Suppose that some valid hidden data is stored with certain public data that has been deleted, the hidden data will finally be relocated to new pages with some other public data. It can be interpreted as that the public data is written to a invalid first write page, just as what happens in dealing with a hidden write request.  
%

Specifically, the hidden data in the victim block (if any) can be encoded and written to empty pages together with certain public data, which comes from either the victim block or a new public write request. As a result, the hidden data that is stored with public data that is subsequently deleted is not lost. 

Moreover, as hidden data are re-encrypted (semantically secure, randomized) during relocation, an adversary cannot link a particular hidden data to a public data it is stored with before and after garbage collection. In effect, deleting public data stored in a particular page does not impact the security and consistency of the hidden data within.

\subsection{I/O Operations}
\label{design:io}

%\blue{
\paragraph{Common Interface for Public and Hidden Data}
As discussed in Section \ref{sec:model}, \sysname~ supports both public and hidden data requests. However, crucially, {\sysname~} does not require different interfaces for accessing public and hidden data. Instead, \sysname~ separates hidden data requests from public data requests by a simple offset convention. Hidden data requests (received through the {\em unchanged FTL interface}) address an offset beyond the physical standard device capacity. This signals to \sysname{} that these requests are addressed to the hidden volume. 

Note that an adversary attempting to access hidden data through this interface would have to provide the correct password at boot-time. Otherwise, the system will be unable to decrypt hidden data. As a result, simply having access to the same  interface does not provide any advantage to the adversary in detecting the presence of hidden data.

%simply observing a single interface leaks nothing about the presence of hidden data to adversaries observing the physical device. 

\paragraph{Preprocessing}
%}
%
%
Upon receiving I/O requests from upper layers, \sysname~ divides the incoming requests into page-level requests first, which are then executed individually.
% it into multiple page requests according to its size (one request may access more than one page). Then \sysname~ performs those page requests one by one.
%
To execute these requests, the first step is a logical-to-physical address translation. \sysname~ first looks up the logical page address in the cached mapping table CMT.
%based on the logical page address for the corresponding mapping information. \todo{don't understand this sentence}
If no hit, either the public or the hidden GTD is queried for the location of the corresponding translation page which contains the target physical page mapping entry which can be used to access the page. The mapping is then also cached in the CMT. If this requires a cache eviction, the least recent used entry will be evicted, resulting an update to its corresponding translation page on the device. And if there is any other mapping entries which belong to the same translation page in the CMT, those entries will be written back to the device simultaneously.
It is important to note that the (either public or hidden) translation pages are written just like actual (public or hidden) data pages. 
% Additionally, the correlated GTD will be updated to keep track of the translation page. 

% To execute a page read request for either public or hidden data, after the address translation, \sysname~ can read and decoded the desired data page accordingly. % you are stating the obivious 

\paragraph{Public Write}
% {\em Public Write.~} 
In the case of a public write, after the address translation, \sysname~ identifies one page for the public data based on the page allocation algorithm in Figure \ref{fig:alloc}. The public data is written to the page following the WOM code based encoding scheme. If the logical address was originally mapped to a valid V1 page (the write request is an update to existing data), the page now transitions to UI1 status and is set to be the {\it Current UI1 Page}. If the logical address was originally mapped to a TI1 page, the page is deleted from the TIQ and then set to be the {\it Current UI1 Page}. 
Finally, the mapping entry is cached in the CMT accordingly. 

\paragraph{Hidden Write}
% {\em Hidden Write.~} 
Hidden page writes require public data to ``cloak'' in: first valid page in the least active public data block. This can then be explained as a simple garbage collection related data relocation. Similarly to the other cases, the corresponding public data mapping information is cached in the CMT. The hidden and the public data are then encoded and written to the {\it Current Empty Page} and the CMT is updated accordingly. Finally, the {\it Current Empty Page} then becomes the next page in the same block, or the first page of a new empty block if the original {\it Current Empty Page} was the last page of a block. The new empty block is selected from the {\it FBL}.

% Note that no page will be pushed into invalid queues as the least active block will be erased soon.
 
\paragraph{TRIM}
% {\em TRIM.~} 
For either a public or a hidden page TRIM request, the deleted data is marked as out-of-date. If the deleted public data was in a V1 page, the page is pushed to the TIQ.
% \todo{can TRIM be issues for both public and hidden data pages?}

\paragraph{Power Loss}
% {\em Power Loss.~} 
Power-loss recovery is not described in DFTL. For simplicity, we assume the physical device comes with standard enterprise grade power loss protection (PLP) backed by capacitors that power up the device for enough time to guarantee caches and other memory resident data structures can be flushed to disk. \sysname~ adds to standard PLP the requirement to write our hidden and public GTDs on power loss also. We also note that if PLP is not available, all data structures can be reconstructed by traversing the entire disk.
% (in the worst case), or in an optimized variant, on the fly during actual operation. 

\paragraph{Encryption}
% 
% Both hidden 
% including the hidden data page and the hidden translation page 
% and public 
{\em Before encoding}, public data and hidden data are first encrypted with different keys using AES-CTR with random IVs. As hidden and public data share physical pages, they can also share the IV in the OOB area of each page.

\section{Security Analysis}
\label{sec:security}

% The security of \sysname~ relies on not only the WOM code that can hide the hidden data stealthily, but also a few other mechanisms used when designing the functions in the FTL. 

The aim of this Section is to show that both the hidden data content and operations are protected from a multi-snapshot on-event adversary. The general idea is that anything that happens between snapshots is a combination of operations. It is then sufficient to show that each such operation does not provide any advantage to a polynomial adversary. 

Specifically, we show that any hidden operation leaves the device in a state indistinguishable from a state resulting from a plausible set of public operations. Then, if all operations are sequential, the effect of any combination thereof (whether or not they include hidden operations) can be explained by a plausible set of public operations. 

\vspace{-0.2cm}
\begin{thm}[]
A computationally-bounded adversary cannot distinguish a physical page containing both public data and hidden data from a page containing only public data written as \secondrecords~. 
% A physical page containing both public data and hidden data is indistinguishable from a page containing only public data written as \secondrecords~. 
% An adversary inspecting an {\sysname~} device cannot recover information about either the contents of hidden data or any evidence of its presence, better than random guessing.
\end{thm}

\vspace{-0.2cm}
\begin{proofsketch}
Lemma \ref{lem0} in Section \ref{sec:pdwom} shows that by construction the 
% WOM code based hidden data 
encoding scheme prevents an adversary from distinguishing a page containing \fullcodes~ (containing hidden data) from a page containing \secondrecords~. 
% Thus, the adversary cannot distinguish whether a page has been written with both public data and hidden data or just been written twice with only public data.
%a page that has been written twice following the WOM code encoding scheme. 

Furthermore, hidden data is encrypted using a semantically-secure randomized cipher. Adversaries can certainly interpret all the pages containing \secondrecords~ as hidden data based on the encode scheme, but all she can get will be the encrypted hidden data, indistinguishable from random. 

Thus, a physical page containing both public data and hidden data is indistinguishable from a page containing only public data written as \secondrecords~.
\end{proofsketch}

% The indistinguishability of the \secondrecord~ and the \fullcode~ prevents the adversary from detecting the hidden data via the physical page content. 
% However, as analyzed in Section \ref{sec:requirement}, a multi-snapshot adversary can observe modifications between snapshots and gain a knowledge of the logical requests. 
% \sysname~ ensures that modifications between snapshots can always be interpreted as the result of operations to public data only. 

% \sysname~ ensures that any operation that involves hidden data is plausibly deniable.
% In other words, the modifications between snapshots can be always interpreted as the result of some operations to public data. 
%In spite of the indistinguishability of the second write pages with the hidden write pages, we can also prove that an  multi-snapshot adversary who can compare the snapshots can also not tell the existence of the hidden data. In other words, every operation that involves a hidden data page is plausibly deniable. The modifications between snapshots can be always interpreted as the result of some operations to only public data. 

% \begin{lemma}[]
% \label{lem2}
% Writing the cmt
% % An adversary inspecting an {\sysname~} device cannot recover information about either the contents of hidden data or any evidence of its presence, better than random guessing.
% \end{lemma}	

\vspace{-0.2cm}
\begin{lemma}[]
\label{lem1}
% A polynomial adversary cannot distinguish an empty page to which hidden data was written (either hidden data or translation pages) from an empty page to which a series of public operations were performed.
% An adversary inspecting an {\sysname~} device cannot recover information about either the contents of hidden data or any evidence of its presence, better than random guessing.

For any page state resulting from writing hidden messages (either hidden data or hidden mapping table) to an empty page, there exists at least one sequence of public operations that results in the exact same state.
\end{lemma}	

\vspace{-0.2cm}
\begin{proofsketch}
Any hidden message $h$ is always written together with some public message $p_2$ (Figure \ref{fig:indistinguish}) to an empty page, resulting in a page state transition from empty to V2.
As shown in Section \ref{sec:requirement}, this page state transition can be plausibly explained as the combination of a sequence of page state transitions (empty $\to$ V1 $\to$ I1 $\to$ V2). Moreover, there exists at least one public operation that can result in each of those page state transitions for the same physical page. 

The page state transition empty $\to$ V1 can be explained by writing some public message $p_1$.  Remember that $p_1$ values are not unique and can be chosen from $2^{(k-1) \cdot x}$ (Section \ref{sec:requirement}) values. The V1 $\to$ I1 transition can be plausibly explained as updating or deleting $p_1$. Finally, recall that $p_2$ was relocated from the block with the least number of valid pages. This transition I1 $\to$ V2 can be plausibly explained as a garbage collection relocation operation.  

% As shown in Section \ref{sec:requirement}, this page state transition (empty $\to$ V2) can be regarded as the combination of a sequence of page state transitions (empty $\to$ V1 $\to$ I1 $\to$ V2) resulting from some public operations. Moreover, as explained in Section \ref{sec:sys}, the hidden data $h$ is not the only possible reason why the public data $p_2$ is written to the empty page. In other words, a plausible set of public data operations will write $p_2$ to the same location.

% Finally, recall that the public data comes from the block with the least number of active public valid pages. This allows writing $p_2$ to be plausibly justified as a garbage collection relocation operation.

% An empty page is written with {\it dual-encoded data} during this write. This can be interpreted as writing the empty page with public data $p_2$ and later invaliding it before the public data $p_1$ is written there as a {\it 2nd write}.
% The indistinguishability between the above two cases is based on the following facts:

% Then this physical page will be interpreted as a {\it 2nd write} page with public data $p_1$, which means the page was written once with public data $p_2$ and later invalided before the public data $p_1$ was written there. 

% \begin{figure}[t!] 
% \centering
% \includegraphics[width=8cm]{update.eps}
% \caption{.}
% \label{fig:update}
% % \vspace{-0.1cm}    
% \end{figure}

The mapping entry for the plausibly appearing public $p_1$ (Figure \ref{fig:indistinguish}) will not be updated on the device until it is evicted from the CMT. Thus, it is highly possible that this mapping entry change does not need to be flushed out to the device (the mapping entry may be updated again before that), as $p_1$ was already out-of-date when $p_2$ was written.

In summary, the results of writing hidden messages to an empty page can be plausibly explained by a series of public operations.
\end{proofsketch}
\vspace{-0.2cm}

\begin{thm}
\label{thm1}
Any page state transition resulting from either a hidden read operation or a hidden trim operation can be plausibly explained by at least one sequence of public operations.
% \todo{is this the same type of statement as the previous theorem? (i understand the proof may be different) -- if so we should use the same language in both theorems}
\end{thm}

\vspace{-0.2cm}
\begin{proofsketch}
As described in Section \ref{sec:sys}, for either a hidden read or a hidden trim, the only possible state change in the device happens when a mapping entry is evicted from the CMT. In this cases, there are two possibilities. (1) the evicted mapping entry is a hidden entry -- in that case a hidden translation page needs to be updated (recall it is treated as a hidden data page) -- and according to Lemma \ref{lem1}, the resulting page state transition can be plausibly explained as the result of a sequence of public operations. Or (2) the evicted mapping entry is a public entry -- in that case a public translation page needs to be updated -- and this can be plausibly explained using public operations only. 
\end{proofsketch}
\vspace{-0.2cm}

% \begin{thm}
% \label{thm2}
% Any page state transition resulting from a hidden trim operation, can also be plausibly explained by at least one sequence of public operations.
% \end{thm}

% \vspace{-0.2cm}
% \begin{proofsketch}
% The proof is identical to the case of the hidden read operation (Theorem \ref{thm1}) since cache eviction is the only possible state change. The page state transition can be plausibly explained as the result of a sequence of public operations.
% % A hidden trim operation deletes some hidden pages, which means that the hidden data in some physical pages will be marked as out-of-data. This only results into a few hidden translation page read as the mapping entries will be loaded to the CMT and updated. However, the deleted pages are not modified until the page is erased and reclaimed during garbage collection. The updated mapping information may be written back to the device later when a cache eviction happens, which results into a hidden page write. As described in Lemma \ref{lem1}, this is indistinguishable from a few public operations. Moreover, since the choice of garbage collection victim block is not related with the status of any hidden data, the access pattern of a hidden trim operation is indistinguishable from a few public operations.
% \end{proofsketch}
% \vspace{-0.2cm}

\begin{thm}
Any page state transition resulting from a hidden write operation, can also be plausibly explained by at least one sequence of public operations.
\label{thm:write}
\end{thm}

\vspace{-0.2cm}
\begin{proofsketch}
A hidden write operation writes the hidden data and updates the corresponding mapping entry. 
The map update happens in the CMT and will be later flushed to the device during a hidden translation page write when a cache eviction happens. 
As proved in Lemma \ref{lem1}, writing either the hidden data or the hidden translation page can be plausibly explained with several public operations. 
%
% Further, any associated hidden map update happens in the CMT and will be later flushed to the device during cache eviction -- also plausibly explained as a result of a sequence of public operations according to Lemma \ref{lem1}.
% As a result, performing a hidden write request can be explained as performing a few public requests.
%
% When writing a hidden page, it will finally presented in the device as a hidden write page, which is indistinguishable from a second write page. Moreover, as shown in Figure \ref{fig:sec}, the first write public page can be interpreted as any public page that is deleted or updated before, which makes the plausible equivalencies non-exhaustive. The missing of the related translation page update can be contributed to the lazy caching flushing scheme as if the mapping information for this virtual public update is later written back in a real translation page write. Similar to a read, the write will also changes the mapping information which will be later updated in the device when the corresponding translation page is written. This can be regarded as a normal hidden page write as illustrated before.
\end{proofsketch}
\vspace{-0.2cm}

%\blue{

\begin{thm}
\label{thm4}
Any page state transitions resulting from a garbage collection operation can be plausibly explained by at least one sequence of public data operations.
% The Garbage collection in \sysname{} does not leak any information about the existence of hidden data.
\end{thm}

\vspace{-0.2cm}
\begin{proofsketch}
This follows by construction. First, note that {\sysname~} select the victim block (the block to be erased) for garbage collection only based on the state of public data in flash devices -- the presence/absence of hidden data has no impact on this selection.  Moreover, page transitions happen only because up-to-date data in the victim block is relocated to new locations. 
% \chen{The states of public page and hidden page is independent from each other}
All data relocations are handled in the same way as new public/hidden data write requests -- {\sysname~} employs the I/O operations discussed in Section \ref{design:io} to complete these requests. Therefore, by leveraging Theorem \ref{thm:write}, we can show that the resulting page state transitions from the hidden write operations performed after a garbage collection can be plausibly explained by a sequence of public operations. 
\end{proofsketch}
\vspace{-0.2cm}

\section{Practical Concerns}
\label{sec:practical}

\paragraph{Crypto Primitive}
To ensure PD, both the public and the hidden data encoded with the WOM code must appear to be indistinguishable from cryptographically secure random data. Thus, before encoding, public data and hidden data are first encrypted (semantically secure, randomized) with different keys in \sysname~. Considering the special application scenario of disk encryption, crypto primitives used in \sysname~ implementation must be chosen carefully. 
For example, when a block cipher mode requiring an initialization vector (IV) is used, each page is usually assigned with a page-specific random IV to enable random access. These IVs must be easily derived from or stored in the storage system. Reusing an IV may result into a catastrophic loss of security.
There are a few special purpose block encryption modes that are specifically designed to securely encrypt sectors of a disk, such as the tweakable narrow-block encryption modes (LRW, XEX, and XTS) and the wide-block encryption modes (CMC and EME). The application of these modes of encryption can prevent attacks such as watermarking, malleability, and copy-and-paste, which is critical for PD as a weak encryption system can amplify an adversary's advantage and enable easy detection of hidden data. 

\paragraph{Storage Capacity}
The use of WOM codes in PEARL amplifies the size of data because a single logical bit is now represented by multiple bits in storage. Therefore, as expected, the overall logical storage capacity (the total amount of logical data that can be written) of the device reduces. We analyse the extent of this reduction in Section \ref{sec:evl}. However, critically, it is worth noting here that logical storage capacity for public data is not impacted by the amount of hidden data stored in the device. In other words, an adversary 
%confusing: 
% can observe the reduction in overall device capacity and infer that {\sysname~} is being used, but 
cannot detect whether hidden data is being stored with any non-negligible advantage.

\paragraph{Wear on Flash Device}
Flash memory has a limited lifetime -- measured as the number of program/erase cycles a block can endure before becoming damaged and unusable. Although erasures are the major contributors to cell wear \cite{jeong2014lifetime}, recent studies show that programming also has a substantial impact on flash cell wear. 
For example, programming MLC cells as SLC \cite{10.1145/2638552} or occasionally relieving cells from programming \cite{179829} can significantly slow down cell degradation, regardless of the number of erasures. Thus, writing a page twice with the WOM code may increase the page wear. In other words, the number of allowed erasures might decrease. 

In \sysname~, a page is written-to (programmed) once in the case of both public data and hidden data being stored there. The page appears to be written-to twice when an adversary observes the device.  This may allow a new type of side channel attack where the adversary estimates page wear to determine the presence of hidden data -- the page wear may end up being slightly decreased than what is expected when the page contains hidden data. A detailed analysis of this physical side-channel and its impact on overall security is the subject of ongoing work. 

%Further research can be conducted regarding the affect of WOM code on the cell wear and the impact on the PD security.
%However, we believe that the difference of the wear should be small and hard to be detected.

\paragraph{Attacks on Weak Passwords}
The security of all PD systems rely on the confidentiality of the hidden encryption key, which is usually derived from a password. There could be several security issues related to passwords, such as online/offline brute force attacks, social engineering, phishing etc. As a first line of defense, \sysname~ requires users to choose strong passwords with high entropy \cite{sec-password} thereby presumably making the system more resilient to these attacks.

% Thus, \sysname~ requires users to choose strong passwords resilient to guessing to protect their encryption keys. There may be online or offline password guessing attacks, especially when weak passwords are used. We assume that users are able to prevent those attacks happening by using the system properly. 

\paragraph{Adequate Public Cover Traffic for Hidden Data}
As in all prior works, \sysname{} requires public data traffic to hide data. Hidden data is written together with public data -- either existing public data relocated during garbage collection or new incoming public data. Garbage collection is triggered only when empty pages are consumed by new public requests. Thus, to enable hidden data writes, sufficient public data traffic is required. In many scenarios this may be a reasonable assumption since in reality the amount of hidden data requiring protection is often less than public data.

\section{Evaluation}
\label{sec:evl}

%\blue{
In this section, we firstly analyse storage and I/O overheads in \sysname{}. We then present a performance evaluation with experimental results. 

\paragraph{Storage Overhead}
The base (3,5) WOM code used (Table \ref{tab:wom_pd} and Section \ref{sec:skew}) in \sysname{} requires 5 physical bits to store 3 bits of public logical data and 1 bit of hidden logical data. Further, storing metadata (e.g. translation pages) requires a few physical blocks. Overall, this reduces the total amount of logical data that can be stored within the total capacity of the device. 

Specifically, the total amount of public data cannot exceed 60\% of the total physical device capacity, while the total amount of hidden data cannot exceed 20\% of the total device capacity. Thus, around 20\% is sacrificed. This can be improved by using a more storage efficient WOM code that supports equal partitions in \sysname~. 
We leave it as future work. 
% \todo{why didn't we do this? need 1 sentence explanation}

%lost due to the use of this particular WOM code. 

%hidden capacity is around 1/5 of the physical capacity (ignoring the metadata). 

%Yet, from an information-theoretic point of view, the sum of the public and hidden capacity cannot be more than the physical capacity of the device since they coexist together in the device simultaneously. 
%\anrin{not sure why the last line is important here?}

\paragraph{I/O Overhead}
Data amplification also contributes to I/O overheads. With the (3,5) WOM code, 5 physical bits encode only 1 hidden bit, which means that 60KB data is accessed from the device for each 12KB of logical hidden data access. Similarly, accessing 12 KB of logical public data requires 20KB of physical data access. This is expected to reduce overall throughput of the system proportionally. 

Moreover, as described in Section \ref{sec:pdwom}, performing a full write or a 2nd write requires reading of some existing public data or the obsolete data in that page. This additional read also contributes to I/O overhead.

Finally, {\sysname~} also requires additional processing time for address translation and data encoding etc., which contributes to I/O overhead as well. 

% \anrin{I did not understand from this point onwards.}
% In addition, the number of physical page operations will increase. This is partially linked with the previous point, but not completely overlapped. The increase of data size may lead to an increase of the number of page operations. For example, if the request data size is much less than the physical page size, then the request may finally need to access only one physical page even if the WOM code is used and the data size is blow up. 
% There are also some additional page operations that is mandtory in \sysname~ for requests of any data size. For example, to write some hidden data, a page read operation is required whatever size the request data is according to our data encoding scheme.
%}

\subsection{Experimental Results}
\label{sec:evl1}

%\blue{
\paragraph{Setup}
\sysname{} was implemented as a core FTL engine in FlashSim \cite{kim2009flashsim}, a popular flash based storage system simulation framework. FlashSim is an event-driven simulator (similar to  DiskSim\cite{disksim}) and is widely used to study the performance implications of different FTL schemes \cite{xu2012cast,zhou2018correlation,hu2011performance,huang2011performance}. 
%It is an event-driven sismulator just like DiskSim\cite{disksim} for disk drive simulation. 
Specifically, the evaluated \sysname{} uses the data encoding scheme based on the (3,5) WOM code as discussed in Section \ref{sec:sys}.
Besides, as a baseline for comparison, DFTL is also implemented and evaluated under the same device settings.
In the experiments below, a 64GB SSD\cite{ssd-para} is simulated and the parameters used for this simulation are listed in Table \ref{tab:para}. The page read, write and erase time are  130 us, 900 us and 10ms, respectively.

% \sysname~ evaluated  
% %5 physical bits are required to store 3 bits of public data and 1 bit of hidden data. 
% Besides, as a baseline for comparison, DFTL is also evaluated under the same settings.

%We use FlashSim \cite{kim2009flashsim} to evaluate the performance of \sysname~.
%% to see how the additional plausible deniability feature affects the storage system performance. 
% FlashSim is a popular flash based storage system simulation framework that is widely used to study the performance implications of different FTL schemes. It is an event-driven simulator just like the DiskSim\cite{disksim} for disk drive simulation. \sysname~ is implemented as the core FTL engine in the simulator. The DFTL is also evaluated under the same setting as the baseline for comparison.
% A 64GB SSD\cite{ssd-para} is simulated whose parameters are listed in Table \ref{tab:para}. The page read, write and erase time are  130 us, 900 us and 10ms, respectively.

% The parameters of NAND flash device simulated come from \cite{ssd-para}, which is a 64GB SSD. The page read, write and erase time are set to be 130 us, 900 us and 10ms respectively as listed in with all the other spec parameters.
%, based on the specification in \cite{gupta2009dftl}.

\begin{table}[!th]
\begin{center}
\begin{small}
\vspace{-0.2cm}
\begin{tabular}{ | c | c | c | c | c |}
\hline
Read & Write & Erase & (Die, Plane, Block, Page) & Page size \\
\hline
130us & 900us & 10ms & (1, 2, 1437, 768) & 16KB \\
\hline
\end{tabular}
\caption{\small{Parameters of the simulated NAND flash device.
\label{tab:para}}}

\end{small}
\end{center}
\end{table}

\paragraph{Logical Volume Capacity}
Although the physical capacity of the SSD simulated is 64GB, the logical capacity for the public volume and the hidden volume are set to 36GB and 12GB respectively. The difference in capacity is due to the use of the (3,5) WOM code, as discussed above. Comparably, the logical volume size when DFTL is used is 54 GB.

%from mainly two reasons. 
%Firstly, the WOM code blows up the size of logical data. For example, as shown in Table \ref{tab:wom_pd}, 5 bits of physical data are stored for every 3 bits of public data. 
%Secondly, the metadata (e.g. translation pages) also consumes a few physical blocks.

% The hidden data capacity in \sysname~ also relies on the specific WOM code used. With the sample (3,5)-WOM code, the hidden capacity is around 1/5 of the physical capacity (ignoring the metadata).
%Using another WOM code that supports equal partition instead of the (3,5) WOM code can lead to another version of \sysname~ where the decrease rate of logical capacity is different.
%Yet, from an information-theoretic point of view, the sum of the public and hidden capacity cannot be more than the physical capacity of the device since they coexist together in the device simultaneously. 

% \anrin{since the reviewers are so concerned with the storage capacity, can this be put in a more visible location? e.g., after you have described the data encoding scheme?}

\paragraph{Initialization} 
FlashSim starts by simulating an empty SSD. However, it is well known that the overall performance of an SSD degrades with increasing logical capacity utilization. Thus, for an accurate evaluation, it is important to start with the device at a state where it has been fairly used for storing and accessing data for all volumes. This requires two things. 
First, the SSD should be ``full'' --  most of the physical pages have been written at least once and contains some data (may be invalid data). Only in this case garbage collection can be triggered. 
Second, the amount of valid data in each volume should be ``equivalent'' relative to volume capacity. In this case, the write amplification due to the relocation of valid data will be comparable. 

Thus, in the initialization phase for the \sysname{} evaluation, the SSD was filled with random data coming from the first halves of the public and the hidden volumes respectively until most of the physical pages have been written once and at least one garbage collection has been invoked. 
When evaluating DFTL, the SSD was filled with random data from the first half of the corresponding logical volume.

\begin{table}[!ht]
\begin{center}
\begin{small}
\vspace{-0.1cm}
\begin{tabular}{ | c | c | c | c | c |}
\hline
 \multirow{2}{*}{Workload} & Avg. Req. & Read & Seq & Avg. Req. Inter- \\
 & Size (KB) & (\%) & (\%) & arrival Time (ms) \\
\hline
Financial1 & 3.47 & 23.2 & 2.0 & 8.19 \\
\hline
%Cello99 & 5.03 & 35.0 & 1.0 & 41.01 \\
Financial2 & 2.45 & 82.3 & 2.0 & 11.08 \\
\hline
Web Search1 & 15.51 & 99.9 & 14.0 & 2.98 \\
\hline
\end{tabular}
\caption{{\small Enterprise-scale workload characteristics.
\label{tab:workloads}}}
%\vspace{8.4pt}
\end{small}
\end{center}
\end{table}

\paragraph{Performance Metrics}
FlashSim reports a total aggregated {\it response time} for each request received. This is a combination of the device service time and the effect of queuing delays. Specifically, the response time not only captures the overhead due to the internal processes in an FTL such as address translation and {\em data encoding}, but also factors in the time spent by the request in I/O queues etc. 
%when the bandwidth is saturated. 

While in certain cases it may be desirable to eliminate scheduling delays etc. from the performance evaluation, this is not possible in the current simulator and would require further kernel instrumentation and more.

\paragraph{Workloads}
We used both real-world workloads and synthetic workloads to benchmark \sysname~ and DFTL. To evaluate performance for real world applications, we used three popular enterprise-scale workload traces (Table \ref{tab:workloads}). This includes two different I/O traces (Financial1 and Financial2) for an OLTP application running at a financial institution \cite{trace-fin}, and an I/O trace from a popular search engine (Web Search1) \cite{trace-web}. 
These traces were particularly selected since (i) their address spaces fit within the capacity of the SSD being simulated, and (ii) they include enough writes to invoke garbage collections. 

Moreover, these traces provide different characteristics which capture numerous real-world usage scenarios. For example, Financial1 is write-dominant while Financial2 and Web Search1 are read-dominant. Further, Web Search1 has more sequential accesses compared to Financial2. Financial1 and Financial2 also have smaller request sizes while Web Search1 requests more data per request on average. The overall parameters for the traces are summarized in Table \ref{tab:workloads}.

%have address spaces that could fit in our SSD size, and include enough write requests to invoke the garbage collector on the device.

In addition to the real-world workloads, we also used synthetic workloads to test throughout of the system under conditions of heavy load.  As shown in Table \ref{tab:workloads}, the average request interval arrival time is usually long in real-world applications. As a result, the available device bandwidth is never fully utilized. In order to test maximum throughput, we ran multiple synthetic workloads where large numbers of requests are submitted to the device at the same time (the request interval arrival time is 0). % This simulates the reads/writes of large files. 
Specifically, 100000 read or write requests are submitted for either public or hidden data, and each of them requests for a data chunk of 16KB. Similarly, the response time is recorded for each request and we calculate the number of request satisfied during each second (IOPS). 
% This gives us the overall throughput of the system.

% Real-world workloads are used to study how the PD feature affects the system performance. We downloaded three popular enterprise-scale workload traces whose characteristics are listed in Table \ref{tab:workloads}. The Financial1 and Financial2 \cite{trace-fin} are both I/O traces from an OLTP application running at a financial institution, while the Web Search1\cite{trace-web} is a read-dominant I/O trace from a popular search engine. It can be seen that the Financial1 is write-dominant while the Financial2 has more reads. The average request size is smaller in the Financial1 and Financial2 when compared to that in the Web Search1 trace.

% Besides, synthetic workloads are also used to test the throughput of the system. A large amount of requests are submitted in a short time to saturate the bandwidth, which simulates the write of a large file (2GB). In this case, the throughput can be derived from the reported response time.

\begin{figure}[t!] 
\centering
\vspace{-0.4cm}  
\includegraphics[width=8cm]{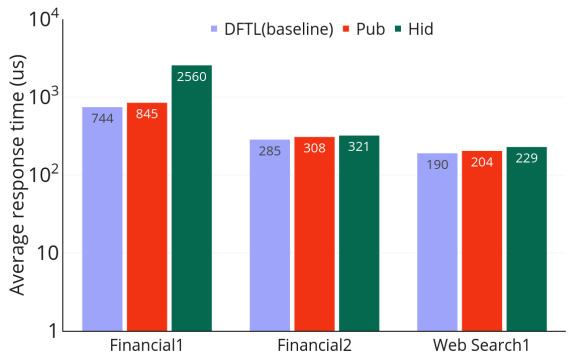}
\vspace{-0.4cm}  
\caption{\small The average response time for three real-world traces with different FTLs (log scale for the y axis, lower is better). }
\label{fig:avg_time}
\vspace{-0.3cm}    
\end{figure}

\paragraph{Results for Application Workloads}
The three workloads listed in Table \ref{tab:workloads} are benchmarked against different FTLs. Figure \ref{fig:avg_time} illustrates average response times for each workload. The y axis is in log scale and the actual values are provided on top of each column for further clarification. 

Generally, I/O requests for hidden data consume more time as compared to public data. Comparing with the baseline (DFTL), the overhead for accessing public data ranges from 6\% to 13\%, while the overhead for accessing hidden data in each workload varies between 13\% to 244\%. The higher overhead for hidden data access is expected since the amplification of data size for hidden data is 3x the amplification of public data. Thus, a hidden data operation requires more physical page accesses compared to a public operation requesting the same data size.

%\chen{}
%\chen{This is mainly because that }
% \anrin{do we need need an intuition here why this is the case?}

% Moreover, hidden data operations are obviously slower than public data operations since the number of bits required to a encode a hidden data bit is much higher than the number of bits to encode public data. As shown in the Table \ref{tab:wom_pd}, 5 physical bits encode only 1 hidden bit, which means that 60KB physical data is accessed in the physical device for each 12KB of hidden data. Comparably, accessing 12 KB of public data accesses only 20KB of physical data.

Further, the average response time increases with increasing percentage of writes in a particular trace. This is more obvious in the case of hidden data accesses. Specifically, for Web Search1, the reported average response time is comparable to the baseline, since more than 99\% of the requests are read requests. On the contrary, the average response time when running Financial1 trace is 2-3x higher than the baseline for hidden data, since most of the requests are writes. 

Specifically, we can conclude that hidden write requests bring much higher overhead than public write requests. This can be explained with the following reasons.
For public data accesses, the overhead for write operations is incurred primarily when the data is written to a 2nd write page. In this case, the old data in the page needs to be read first. A similar overhead is incurred during hidden writes -- the public data that will be stored along with the hidden data needs to be read first. However, as hidden data has a larger amplification due to the WOM code compared to public data, the hidden data may be spread across more pages. And each page of hidden data requires a page of public data to be read. As a result, hidden data writes usually require more page operations than public data writes. 
In addition, a hidden write also requires updating the map entry for the corresponding public data. This may result in additional page accesses. 
Thus, overall, the overheads for hidden writes are much higher than the overheads for public writes.

Interestingly, the above results indicate that the additional page operations are the main contributors to performance overhead rather than additional data processing (data encoding etc).

\begin{figure}[t!] 
\centering
\vspace{-0.4cm} 
\includegraphics[width=8cm]{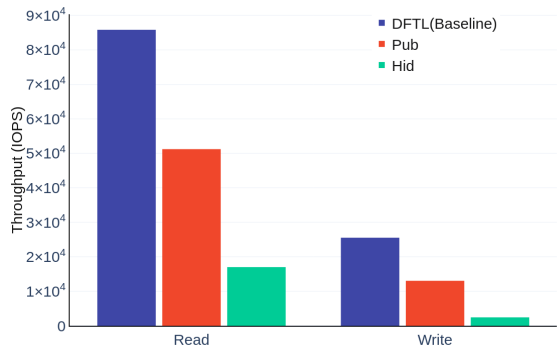}
\vspace{-0.4cm}  
\caption{\small Throughput comparison between DFTL (baseline) and \sysname~. \sysname~ is slower mainly due to data amplification resulting from the use of the WOM code.}
\label{fig:throughput}
\vspace{-0.3cm}    
\end{figure}

\paragraph{Results for Synthetic Workloads}
%
% We test the throughput of different FTLs by running our synthetic workload in the simulator. 
%
%\anrin{I think this has been defined before and we do not need to reiterate the process here.}
%As all the requests are submitted at the same time, we calculate the number of requests satisfied in each one second interval -- this gives the throughput in IOPS for the system. Throughput results in IOPS are presented in Figure \ref{fig:throughput}
%
%The result shows that the IOPS calculated is stable after the first response.
%
The baseline (DFTL) read throughput is around $8.5*10^4$ IOPS and write throughput is around $2.5*10^4$ IOPS.  \sysname~ public throughput is around $5*10^4$ IOPS for reads and $1.3*10^4$ IOPS for writes, while the hidden throughput is $1.7*10^4$ IOPS for reads and $2.4*10^3$ IOPS for writes. In other words, \sysname{} throughput is around 60\% of the baseline for public data, and 10\% -- 20\% of the baseline for the hidden data.  

The performance penalty for public data operations is primarily due to the data amplification resulting from the WOM code: 5 physical bits are used to represent only 3 bits of public data. Meanwhile, additional page reads required during the {\it 2nd write} also reduces the write throughput of public data. 

Write amplification due to WOM codes also significantly affects the throughput for hidden data operations since 5 physical bits are required for 1 hidden bit. Besides, additional page reads and writes are required for public data that is written together with hidden data and plausibly explains the changes to the device. This explains why the hidden write throughput is around 10\% instead of 20\% of the baseline.

\section{Conclusion}

% {\sysname~} is a flash file system that is ``invisible'' (device layouts identical with that of a standard file system), provides redundancy, handles overwrites, survives data loss, and is secure in the presence of multi-snapshot adversaries. 
% {\sysname~} is efficient and its public data operations are less than 15\% slower than standard YAFFS. 
% Hidden data operation throughputs are of the same order of magnitude as that in existing plausible deniability systems secure against multi-snapshot adversaries.

{\it \sysname{} is the first system that achieves strong plausible deniability for NAND flash devices, secure against realistic multi-snapshot adversaries.}

\sysname~ is based on a new data encoding scheme base on specially designed WOM codes -- the first scheme that allows hidden data to surreptitiously coexist in the same physical page as public data. 
% Based on the special hidden data encoding scheme, we then designed a plausibly deniable FTL \sysname~. It securely processes the I/O request from the upper layers, manages the unavoidable inherent mappings from logical to physical pages and reclaims pages occupied by the obsolete data. More importantly, 
By enabling plausible explanations for all state transitions base on public operations only, \sysname~ ensures that an on-event multi-snapshot adversary cannot detect the existence of hidden data.
\sysname~ performance is practical and real-world workloads perform comparably with the case of running on a standard device without plausible deniability assurances.

\bibliographystyle{plain}
\bibliography{MyBib}

\end{document}

%% file: Macros.tex
\usepackage[]{hyperref}
\hypersetup{
	linktoc=all, 
	hidelinks, 
	colorlinks=true,
	allcolors=blue
}

\usepackage{balance}
\usepackage{subfigure}
\usepackage{epsfig}
\usepackage{graphicx}
\usepackage{amsmath}
\usepackage{amssymb}
\usepackage{amsfonts}
\usepackage{verbatim}
\usepackage{graphicx,url}
\usepackage{color}
\usepackage{wrapfig}

\usepackage{framed}
\usepackage{algorithmic}
\usepackage{setspace}

\usepackage{amsthm} 
\usepackage{changes}
\usepackage[normalem]{ulem}
\usepackage{pgfplots}
\usepackage{multirow}

\newenvironment{proofsketch}{\par\noindent\textsl{Proof (sketch):}\kern1ex}%
{\hfil\hskip2em\penalty250\parfillskip=0pt\finalhyphendemerits=0$\qed$%
\par\smallskip\par}

{\smallskip \list{}{\leftmargin=0.2in\rightmargin=0.2in}\item[] \em}%
{\endlist}

{\hfil\hskip2em\penalty250\parfillskip=0pt\finalhyphendemerits=0$\qed$%
\par\smallskip\par}

\newtheorem{definition}{Definition}

\newtheorem{thm}{Theorem}
\newtheorem{requirement}{Requirement}

\newtheorem{lemma}{Lemma}
\newcommand{\sysname}[1]{PEARL}{\ignorespaces}
\def \etal {et al.~}
\renewcommand{\paragraph}[1]{\noindent {\bf #1.}~}

\newcommand{\womrecord}[1]{{\it WOM write codeword}}{\ignorespaces}
\newcommand{\womrecords}[1]{{\it WOM write codewords}}{\ignorespaces}
\newcommand{\firstrecord}[1]{{\it 1st WOM write codeword}}{\ignorespaces}
\newcommand{\secondrecord}[1]{{\it 2nd WOM write codeword}}{\ignorespaces}
\newcommand{\firstrecords}[1]{{\it 1st WOM write codewords}}{\ignorespaces}
\newcommand{\secondrecords}[1]{{\it 2nd WOM write codewords}}{\ignorespaces}
\newcommand{\fullcode}[1]{{\it full write codeword}}{\ignorespaces}
\newcommand{\fullcodes}[1]{{\it full write codewords}}{\ignorespaces}

\newcommand{\prt}{\ensuremath{\mathsf{prt}}}
\newcommand{\Set}[1]{\ensuremath{\{#1\}}}
\newcommand{\w}{\ensuremath{\mathsf{w}}}
\newcommand{\enc}{\ensuremath{\mathcal{E}}}
\newcommand{\dec}{\ensuremath{\mathcal{D}}}

\usepackage{tikz}

\newlength{\bibitemsep}
\newlength{\bibparskip}
\let\oldthebibliography\thebibliography
\renewcommand\thebibliography[1]{%
  \oldthebibliography{#1}%
  \setlength{\parskip}{\bibitemsep}%
  \setlength{\itemsep}{\bibparskip}%
}